%% file: Concept_for_an_Automatic_Annotation_of_Automotive_Radar_Data_Using_AI-segmented_Aerial_Camera_Images.tex
\documentclass[conference]{IEEEtran}
\IEEEoverridecommandlockouts
\usepackage{cite}
\usepackage{amsmath,amssymb,amsfonts}
\usepackage{algorithmic}
\usepackage{graphicx}
\usepackage{textcomp}
\usepackage{xcolor}
\usepackage{makecell}
\usepackage{soul}

\usepackage{tikz}
\usepackage{pgfplots}
\usepackage{hyperref}

\usepackage{subcaption} 
\usepackage{siunitx}
\usepackage{tabularx}
\def\BibTeX{{\rm B\kern-.05em{\sc i\kern-.025em b}\kern-.08em
    T\kern-.1667em\lower.7ex\hbox{E}\kern-.125emX}}

\makeatother
\begingroup
\catcode`\_=13
\gdef_#1{\sb{\mathrm{#1}}}
\endgroup
\newcommand\enableuprightsubscripts{\catcode`\_=12\relax}

\makeatletter
\enableuprightsubscripts       

\newcommand\copyrighttext{%
  \footnotesize \textcopyright 2023 IEEE. Personal use of this material is permitted. Permission from IEEE must be obtained for all other uses, in any current or future media, including reprinting/republishing this material for advertising or promotional purposes, creating new collective works, for resale or redistribution to servers or lists, or reuse of any copyrighted component of this work in other works.}
\newcommand\copyrightnotice{%
\begin{tikzpicture}[remember picture,overlay]
\node[anchor=south,yshift=10pt] at (current page.south) {\fbox{\parbox{\dimexpr\textwidth-\fboxsep-\fboxrule\relax}{\copyrighttext}}};
\end{tikzpicture}%
}

\begin{document}

\title{Concept for an Automatic Annotation \\ of Automotive Radar Data Using \\ AI-segmented Aerial Camera Images}

\author{\IEEEauthorblockN{Marcel Hoffmann$^{1}$, Sandro Braun$^{1}$, Oliver Sura$^{1}$, Michael Stelzig$^{1}$, Christian Schüßler$^{1}$,\\ Knut Graichen$^{2}$, Martin Vossiek$^{1}$}
\IEEEauthorblockA{\textit{$^{1}$Institute of Microwaves and Photonics (LHFT) \quad $^{2}$Chair of Automatic Control (LRT)} \\
\textit{Friedrich-Alexander-Universität Erlangen-Nürnberg (FAU)}\\
Erlangen, Germany \\
\{marcel.mh.hoffmann, sandro.braun, oliver.sura, michael.stelzig,
christian.schuessler,\\ knut.graichen, martin.vossiek\}@fau.de}
}

\maketitle
\copyrightnotice
\vspace{-0.2cm}

\begin{abstract}
This paper presents an approach to automatically annotate automotive radar data with AI-segmented aerial camera images.
For this, the images of an unmanned aerial vehicle (UAV) above a radar vehicle are panoptically segmented and mapped in the ground plane onto the radar images.
The detected instances and segments in the camera image can then be applied directly as labels for the radar data.
Owing to the advantageous bird's eye position, the UAV camera does not suffer from optical occlusion and is capable of creating annotations within the complete field of view of the radar.
The effectiveness and scalability are demonstrated in  measurements, where 589 pedestrians in the radar data were automatically labeled within 2 minutes.
\end{abstract}


\section{Introduction}

In the field of autonomous driving, radar systems are one of the key components for generating environment maps and detecting vulnerable road users (VRU) \cite{b1}.
To interpret the measured data, machine learning approaches are increasingly being used, for example, to classify pedestrians \cite{b2}.
For a better understanding of the environment, radar maps can also be semantically segmented using neural networks (NNs) \cite{b3}.

However, a major challenge is the acquisition of large, well-labeled datasets required for the training of an artificial intelligence (AI).
This is because manual annotation at each stage of radar processing is nontrivial and time consuming.
Especially due to the often poor angular resolution, labeling without high-quality reference is almost impossible.

To overcome this challenge, automated methods to annotate radar data are required.
Camera images are commonly used for this task, from which labels for the radar data are directly extracted using existing, well-trained NNs for image interpretation.
In \cite{b4}, objects in camera images are marked with bounding boxes to create automatic labels for range-Doppler (RD) maps from frequency-modulated continuous-wave (FMCW) radar data.
One step further along the FMCW processing chain (cf. Section II.), the authors in \cite{b5} label the range-Doppler-azimuth (RDA) cubes using camera images with a pixel-correct instance segmentation to train a NN for the detection of VRUs.
Radar point clouds, at the end of the FMCW process chain can also be automatically labeled using panoptically segmented camera images as a profitable combination of instance segmentation and semantic segmentation \cite{b6}.
Given the orientation of the camera in the vehicle, it is difficult to map camera images to the typically horizontal radar planes.
This is solved in \cite{b7} with a multi-sensor labeling process that additionally uses lidar to spatially synchronize the RD maps with the camera images with higher accuracy.

\begin{figure}[t]
    \centering
    \vspace{-0.2cm}
    \includegraphics[width=0.98\linewidth]{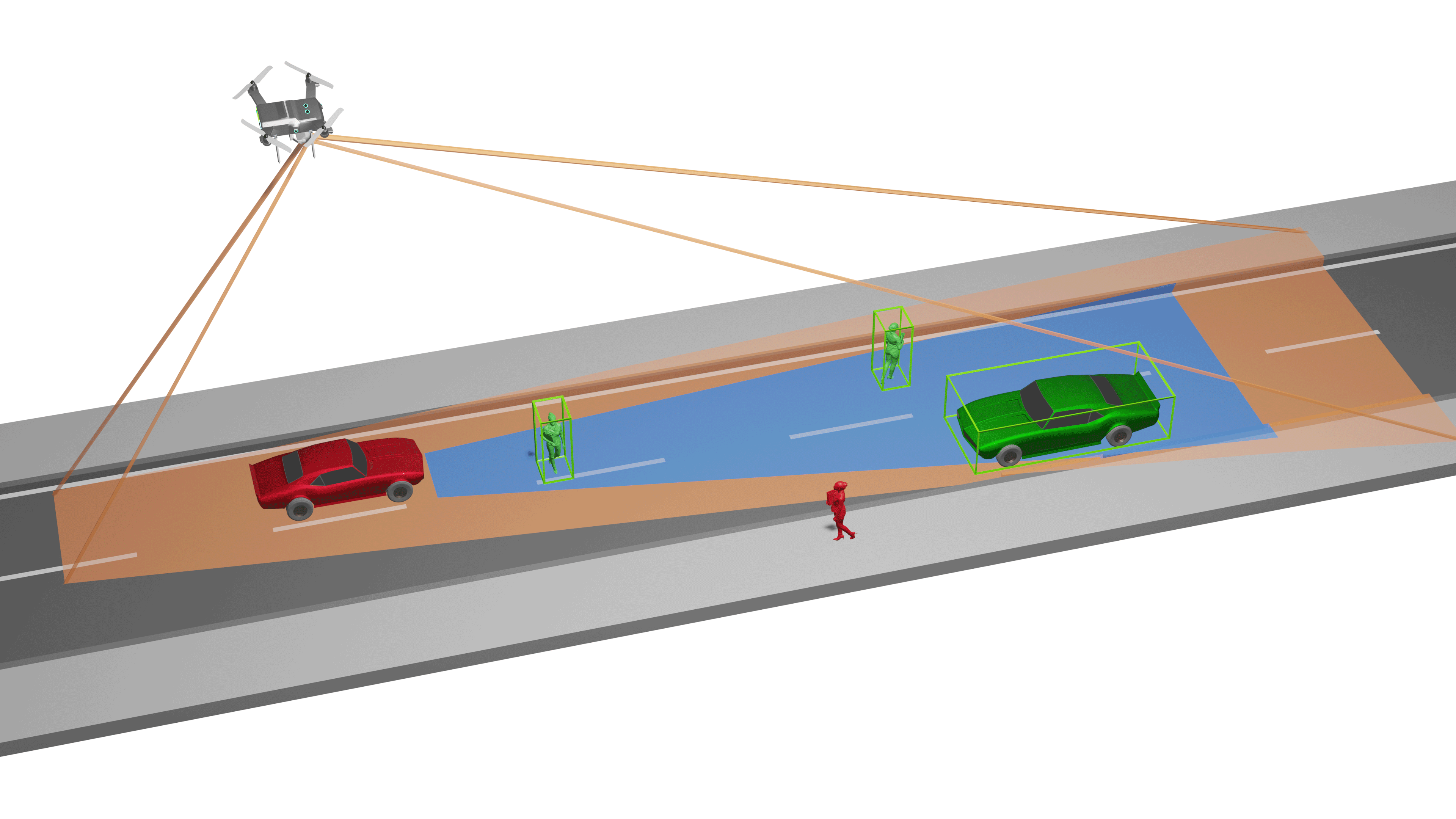}
    \vspace{-0.3cm}
    \caption{Automatic annotation of radar data from a vehicle (red) with labels (green bounding boxes) taken from segmented, synchronized aerial images. With this setup, the FoV of the UAV's camera (orange) covers the complete radar FoV (blue).}
    \label{fig:concept}
\end{figure}

\begin{figure*}[tb]
    \centering
    \includegraphics[width=170mm]{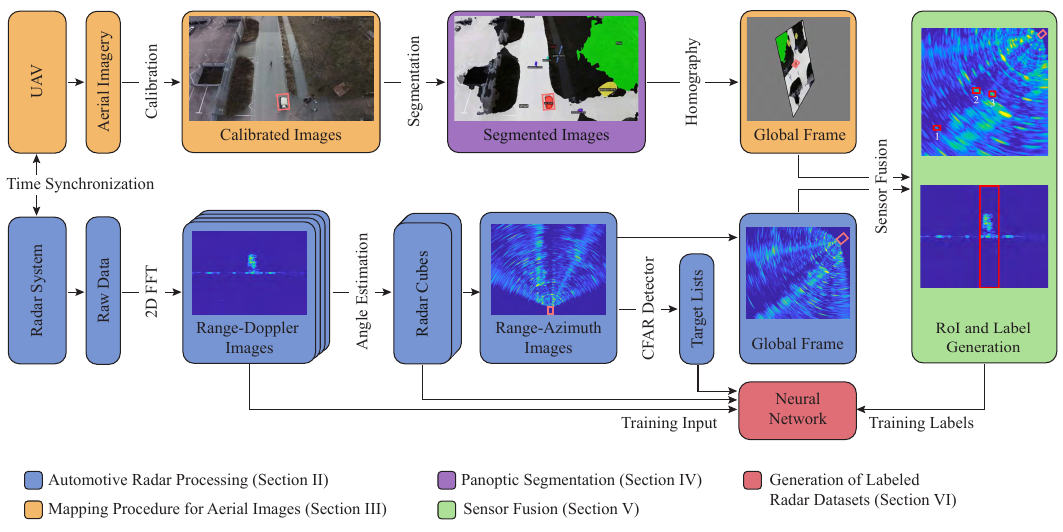}
    \vspace{-0.2cm}
    \caption{Overview of the proposed approach for an automatic annotation of automotive radar data. The top row describes the processing of the UAV images, while the lower row corresponds to the radar processing. Colors match the sections as stated.}
    \label{fig:process_chain}
\end{figure*}

However, for these methods, labels can only be generated in areas that can be detected by the reference sensors from the vehicle’s perspective.
Furthermore, camera and lidar are notably affected by occlusion and, therefore, do not cover the complete radar field of view (FoV).
As a result, the unique benefits of radar can very often not be satisfactorily captured and labeled.
This includes scenes in which several vehicles drive in front of each other.
While radars can image this correctly, cameras and lidars only see the first car.

In \cite{b8}, a camera mounted on an unmanned aerial vehicle (UAV) is used to capture a larger portion of the radar FoV to determine the perception errors of a sensor on the vehicle and thus obtain a generally applicable sensor model.

Based on the perspective gain depicted in Fig.~\ref{fig:concept}, we present an approach to use a spatiotemporally synchronized UAV with camera that is positioned above the vehicle.
With this setup, we automatically generate labels for radar data to train arbitrary NNs.
The paper is structured according to our developed process chain shown in Fig.~\ref{fig:process_chain}.
In Section~II, the processing of radar data is discussed (blue in Fig.~\ref{fig:process_chain}).
Section~III proposes a procedure to map aerial images to the radar plane (orange in Fig.~\ref{fig:process_chain}).
The deployed panoptic segmentation for the interpretation of UAV camera images is presented in Section~IV (purple in Fig.~\ref{fig:process_chain}).
In Section~V, the fusion of radar data and segmented camera images is performed to extract regions of interest (RoI) from the radar images based on the camera labels (green in Fig.~\ref{fig:process_chain}).
Finally, Section~VI describes methods to use the fused data to automatically generate labeled radar datasets for the training of arbitrary NNs (red in Fig.~\ref{fig:process_chain}).
The conclusion is presented in Section~VII.

\section{Automotive Radar Processing}

Automotive radar applications are often based on radars transmitting sequences of linear FMCW signals that can be processed with techniques presented in \cite{b9}.
Our radar processing chain for the proposed automatic annotation is depicted in blue in Fig.~\ref{fig:process_chain}.
It is based on data gathered with the \SI{77}{\giga\hertz} multiple-input multiple-output (MIMO) FMCW radar AVR-QDM-110 from Symeo GmbH, an indie Semiconductor company, that is mounted to a test vehicle.
In Fig.~\ref{fig:process_chain}, this vehicle is always highlighted by a pink rectangle.
In the first step, the RD maps are generated from the raw data using a 2D fast Fourier transform (FFT) with suitable zero-padding and windowing.
For this, 3 out of 12 transmit (TX) and all 16 receive (RX) antennas are used to form a uniform linear array (ULA) with 48 elements and an unambiguous horizontal FoV of approximately \SI{140}{\degree}.
Each sequence consisted of \num{128} chirps per TX antenna with a bandwidth of \SI{1}{\giga\hertz}.

In the second step, the azimuth angle is estimated by an FFT along the calibrated ULA to create the 3D RDA cube.
Here, beamformers or even super-resolution approaches can alternatively be used for the angle estimation \cite{b9}.
Subsequently, a range-azimuth (RA) image representing the horizontal radar plane can be derived.
In the final step, the RA image is transformed to Cartesian coordinates and then aligned with the global frame for a simplified fusion with the segmented camera image, which will be described in Section~V.

\section{Mapping Procedure for Aerial Images}

\begin{figure*}[tb]
    \centering
    \subfloat[]{
        \includegraphics[width=0.2\linewidth]{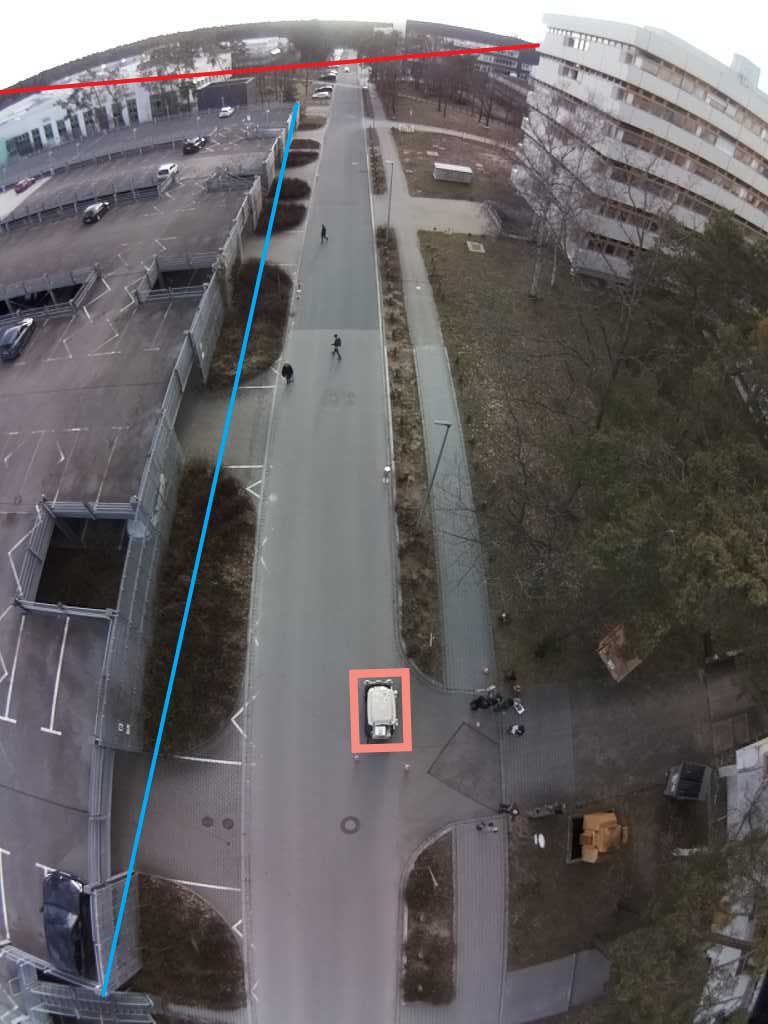}}
   \hspace{0.5cm}
	\subfloat[]{
        \includegraphics[width=0.2\linewidth]{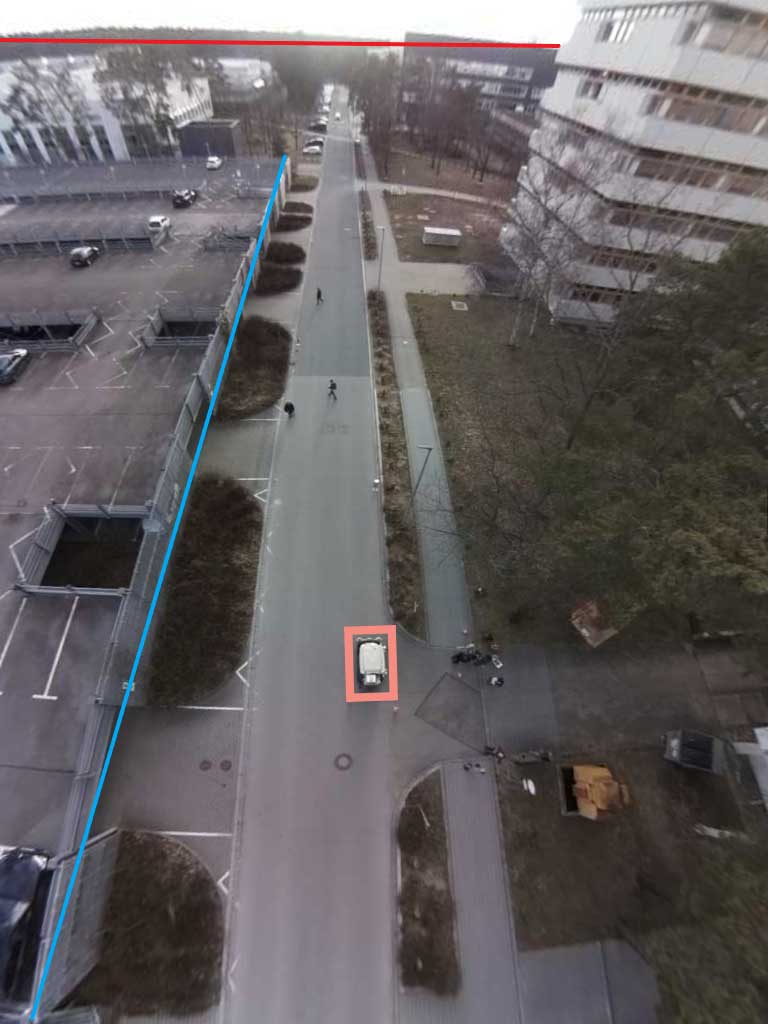}}
  \hspace{0.5cm}
	\subfloat[]{
        \includegraphics[width=0.2\linewidth]{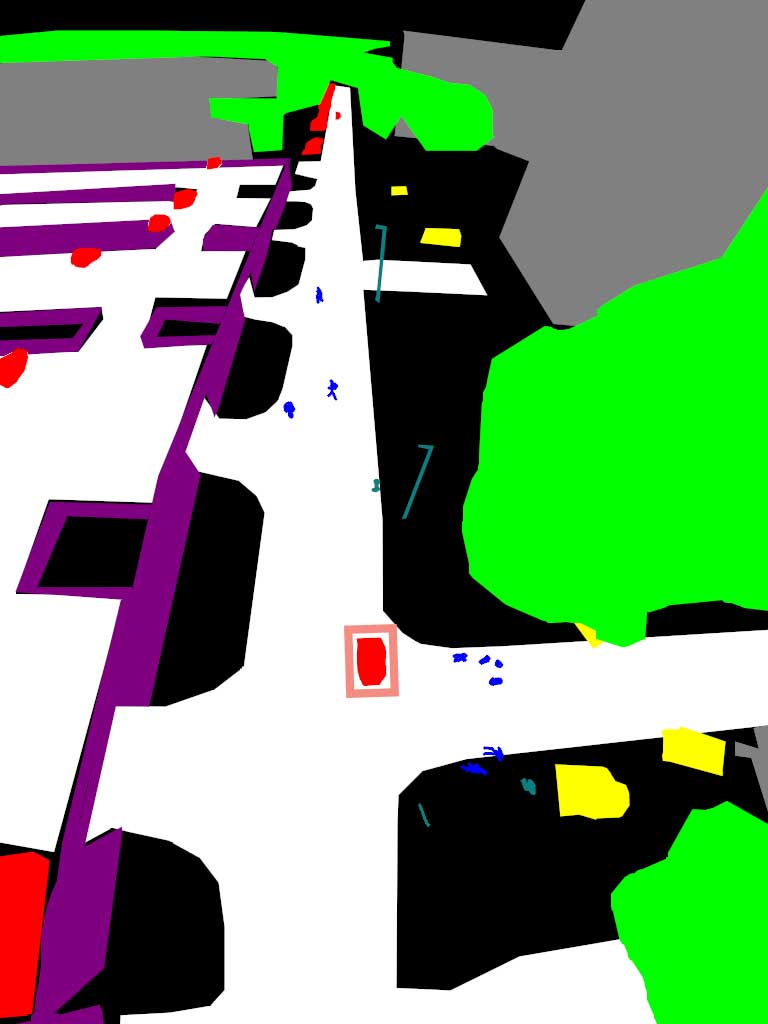}}
  \hspace{0.5cm}
	\subfloat[]{
        \includegraphics[width=0.2\linewidth]{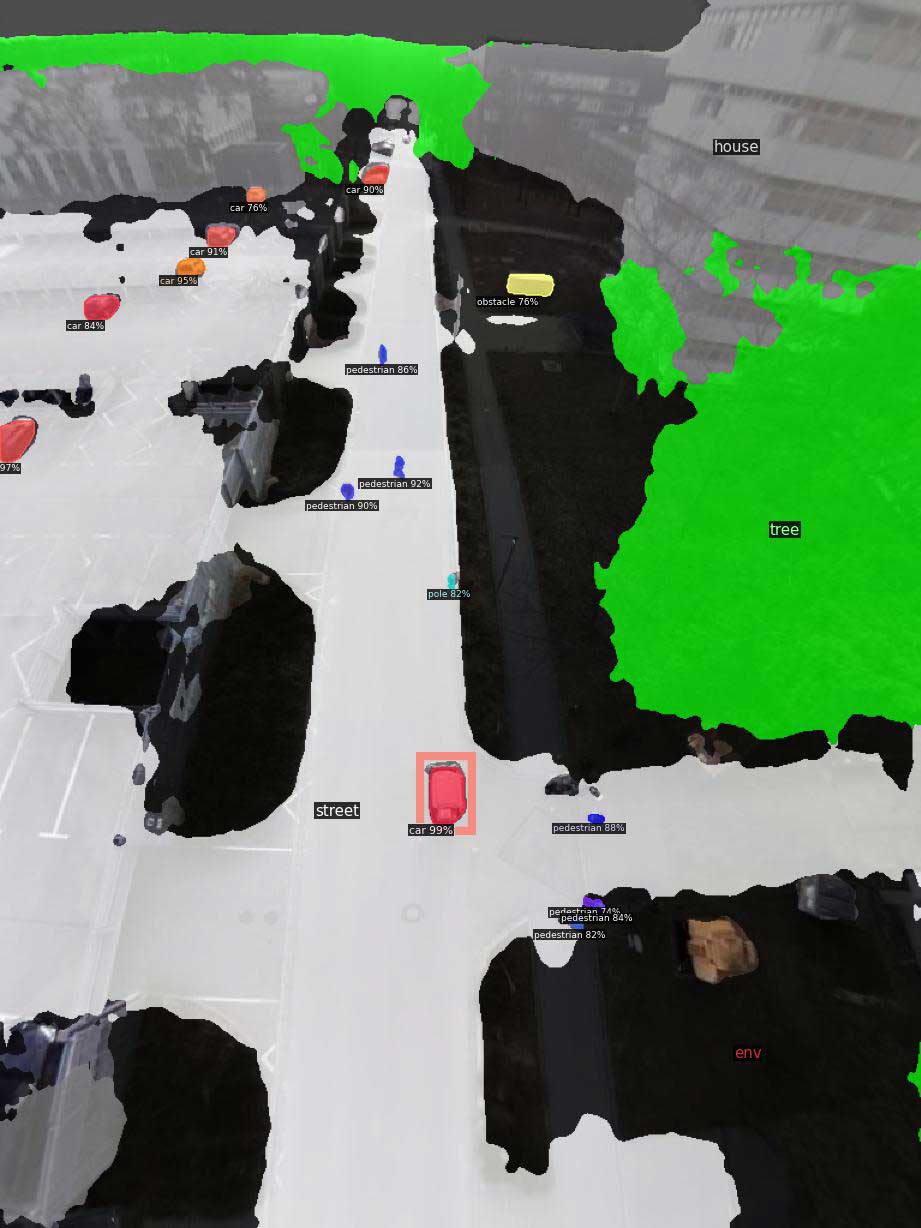}} 
     \vspace{-0.2cm}   
    \caption{Processing chain of the UAV images. (a) shows the original 
 image, (b) presents the undistorted image after calibration, (c) demonstrates the labelling process for training, and (d) presents the panoptically segmented image. In (c) and (d), black, white, green, red, yellow, and blue correspond to the environment, street, vegetation, cars, obstacles, and pedestrians, respectively. The radar vehicle is marked with a pink rectangle in all four images. Three pedestrians are walking within the radar FoV.}
    \label{fig:img_chain}
\end{figure*}

A crucial part of labeling radar data is the precise mapping of the aerial images to the horizontal radar plane that has been transformed to the global frame in the previous section.
All steps concerned are highlighted in orange in Fig.~\ref{fig:process_chain}.

For correct mapping, the camera images have to be calibrated to remove distortions caused by the lens.
Next, the observed ground area has to be estimated based on the UAV's position.
Finally, the aerial image plane has to be transformed to the ground plane using the homography matrix of the two planes.
As the observed area in our setup is small compared to other remote sensing applications, we consider the earth to be a horizontal plane.
However, for large areas, distortions caused by terrain or sensor motion have to be corrected.
This procedure of precisely converting images to a map-suitable form is called orthorectification.
\cite{b10} provides an overview of methods for generating UAV orthoimages.

\subsection{Camera Calibration}

We used the Raspberry Pi Camera Module V2.1 with a lens with a physical vertical FoV of $\alpha = \SI{115}{\degree}$ and a horizontal FoV of $\beta = \SI{80}{\degree}$.
The UAV (Holybro S500 V2) was positioned at a height of \SI{25}{\m} and was connected to the vehicle's Robot Operating System via WLAN.
The camera was tilted to the ground at an angle of \SI{60}{\degree}.
With this setup, it is possible to overlook the whole scene and create a large overlap with the radar's FoV without the need for large altitudes.
Owing to the lens characteristics, a strong radial and tangential distortion occurred in the images, as depicted in Fig.~\ref{fig:img_chain}a).
Assuming a pinhole camera model, the distortion can be corrected by a calibration method described in \cite{b11}, which we realized with functions from the OpenCV software library \cite{b12}.

The calibrated images exhibit significantly reduced distortion, thus depicting the real geometry more accurately.
This is also visible from the colored lines in the original and calibrated example images in Fig.~\ref{fig:img_chain}a) and Fig.~\ref{fig:img_chain}b), respectively. 

\subsection{Determination of the Observed Ground Area}

To establish a relationship between the global frame and the image plane, the position, orientation, and FoV of the camera must first be determined with respect to the world coordinate system.
The position of the center point $c_0$ of the observed ground plane can be determined via the rotation and translation of the camera with respect to the global coordinate system.
For this, the data gathered from the drone's flight controller (Holybro Pixhawk 4) were used.
It features a Global Navigation Satellite System (GNSS) module and an Attitude And Heading Reference System (AHRS) for pose estimation.

Based on the pinhole camera model, the observed ground area can be described as a quadrilateral using four vectors from each image corner to the corners of the ground plane \cite{b13}.
The four corner coordinates $c_{T/B, L/R}$ (top/bottom, left/right) or vertices of the observed ground area can be described by 
\begin{equation}
    c_{T/B,L/R} = \left(\ell_{T/B}, \ \ell_{L/R}\right)^\text{T}\text{,} \\
 \end{equation}
with $\ell$ being the distance from $c_0$ in the respective dimension:
\begin{equation}
    \begin{split}
        \ell_{T,B} &= z\cdot \tan\left(\pm\frac{\alpha}{2} \pm \tan^{-1}\left(\frac{h}{2f}\right)\right)\text{,} \\
        \ell_{L,R} &= z\cdot \tan\left(\pm\frac{\beta}{2} \pm \tan^{-1}\left(\frac{w}{2f}\right)\right)\text{.}
    \end{split}
\end{equation}
Here, $z$ denotes the UAV's altitude, and $h$, $w$, and $f$ are lens height, lens width, and focal length, respectively \cite{b13}.

\subsection{Homography}

In the final step, the homography approach aimed to realize a projection of the camera image to the global frame (cf. Fig. \ref{fig:concept})\cite{b14}\cite{b15}.
For this, the four vertices of the calibrated camera image were mapped to the previously computed corner coordinates $c_{T/B, L/R}$ on the horizontal ground plane.
As a result, the camera image was also mapped to the global frame, which is depicted in Fig.~\ref{fig:process_chain} in orange and blue, respectively.
Consequently, objects in both radar and camera images overlapped, and it became possible to fuse these two data types, which will be described in Section V (green in Fig.~\ref{fig:process_chain}).

\section{Panoptic Segmentation of Aerial Images}

To automatically generate labels and extract information from camera images, the  images are interpreted using a NN.
For this, we chose a panoptic segmentation because it combines the advantages of instance segmentation and semantic segmentation and, consequently, offers the best flexibility for the actual task of labeling the radar data.
Therefore, it allows the annotation of individual objects, such as for a radar-based VRU classification.
Moreover, uncountable regions like the street can be segmented pixel-correct, which could be used to generate a training dataset for the segmentation of radar gridmaps.
In Fig.~\ref{fig:process_chain}, this step is colored purple.

\subsection{Manual Annotation of Camera Images}

In this application, the aim of panoptic segmentation was to label radar data optimally rather than extract the maximum information content from the image.
Therefore, we defined a small number of 12 countable and non-countable classes that are of specific relevance in the radar context because of their unique features in radar images.
The classes listed in Table~\ref{tab:classes} contain moving road users like pedestrians and cars as well as static objects like poles appearing as point targets.

Based on these classes, we manually annotated a small and specific dataset to train the panoptic segmentation of the UAV images. 
Due to the easy interpretability of camera images, it is possible to annotate images efficiently and quickly even without a technical background, which, again, is not feasible for radar images.
For the annotation process, image segments were marked with 12 colors representing the 12 classes.
An example for this is shown in Fig. \ref{fig:img_chain}c).

\begin{table}[tb]
\centering
\caption{Classes for panoptic segmentation.}
    \begin{subtable}{\linewidth}  
    \centering  
        \begin{tabular}{c|c}
        
        \hline
        \textbf{Class} & \textbf{Description} \\
        \Xhline{4\arrayrulewidth}
        Environment & Background (e.g., small vegetation/objects, sidewalks) \\
        \hline
        Street & Roads \\
        \hline
        Trees & Larger vegetation with trunk  \\
        \hline 
        Houses & Buildings with walls  \\
        \hline
        Barriers & Continuous wall-structures not resembling houses\\
        \hline
        Poles & Cylindrical, point-target-like objects \\
        \hline
        Obstacles & Free-standing, large obstacles (e.g., power boxes, stones) \\
        \hline
        Cars & Moving and parking cars \\
        \hline
        Trucks & Larger trucks and buses (bigger than cars) \\
        \hline
        Motorbikes & Motorized two-wheeler including riders \\
        \hline
        Bikes & Bicycles including riders \\
        \hline
        Pedestrians & Moving and static pedestrians\\
        \end{tabular}
    \end{subtable}
\label{tab:classes}

\end{table}

\subsection{Training of the Neural Network for Camera Images}

For this application, the Panoptic Feature Pyramid Network "PanopticFPN" model from \cite{b16} was chosen, which uses the open-source platform Detectron2 from Meta Research \cite{b17}.

The final training dataset contained \num{528} images in total; \num{338} of which were manually annotated images and \num{190} images were taken from the UAVid dataset \cite{b18}.
The validation dataset contained 96 manually annotated images.
The distribution of the instances of each class are listed in Table~\ref{tab:instances_dist}.
The training was performed on an Nvidia GeForce RTX 3090 graphics card with \SI{24}{\giga\byte} VRAM and CUDA v11.3.
The batch size was set to 6, and the number of iterations were \num{15000} with a learning rate reduction to \SI{10}{\percent} at \SI{60}{\percent} and to \SI{1}{\percent} of the original learning rate at \SI{85}{\percent} of the total iteration number.

\begin{table}[tb]
	\centering
 	\caption{Distribution of countable class instances.}
	\begin{subtable}[][][c]{0.46\linewidth}
		\captionsetup{justification=centering}
		\centering
        \caption{Training dataset.}
		\begin{tabular}{c|S[table-format=4]}
			\hline
			\textbf{Class} & \textbf{Instances} \\
			\hline
			Barriers & 1452 \\
			Poles &  5745\\
			Obstacles &  2943\\
			Cars &  14280\\
			Trucks &  94\\
			Motorbikes &  79\\
			Bikes &  1118\\
			Pedestrians & 4520 \\
			\hline
			\textbf{Total} & 30231 \\
			\hline
		\end{tabular}
	\end{subtable}
	\begin{subtable}[][][c]{0.46\linewidth}
		\centering
        \caption{Validation dataset.}
		\captionsetup{justification=centering}
		\begin{tabular}{c|S[table-format=4]}
			\hline
			\textbf{Class} & \textbf{Instances} \\
			\hline
			Barriers & 646 \\
			Poles &  1976\\
			Obstacles &  1035\\
			Cars &  844\\
			Trucks &  3\\
			Motorbikes &  20\\
			Bikes &  354\\
			Pedestrians & 222 \\
			\hline
			\textbf{Total} & 5100 \\
			\hline
		\end{tabular}
	\end{subtable}
	\label{tab:instances_dist}
\end{table}

\subsection{Evaluation of the Trained Model}

An exemplary result of panoptic segmentation is shown in Fig.~\ref{fig:img_chain}d).
To avoid falsification, the corresponding image in Fig.~\ref{fig:img_chain}c) was not included in the training.

In the following evaluation, we set the focus on pedestrians because VRU classification has high relevance in radar applications.
This is a suitable verification of the proposed approach, because on the one hand, the moving pattern of a pedestrian can be easily detected as a Doppler signature in RD maps.
On the other hand, they are harder to detect from the UAV's perspective due to their small size, which also requires a precise coregistration of camera and radar.

\begin{table}[tb]
    \centering
    \caption{Metrics for the trained model.}
    \begin{tabular}{c|c|S}
         \textbf{Metric} &  \textbf{IoU interval (in \%)} & \textbf{Value (in \%)}\\
         \hline
         $mAP$ &  50:5:95 &     6.89 \\
         $mAP$-50 &  50 &  14.53 \\
         $mAP$-75 &  75 &  6.22 \\
         $AP_{car}$ & 50:5:95 &   25.70 \\
         $AP_{bike}$ & 50:5:95 &  5.55 \\
        $AP_{pedestrians}$ &  50:5:95 & 17.67\\
    \end{tabular}
    \label{tab:metrics_panseg}
\end{table}

To evaluate the accuracy of the detections, the intersection over union (IoU) will be calculated as,
\begin{equation}
	IoU = \frac{|T \cap P|}{|T \cup P|} \text{,}
\end{equation}
where $T$ denotes the area of the ground-truth and $P$ is the area of the object proposed by the panoptic segmentation.
If the IoU is larger than a threshold $k$, then a detection is considered a true-positive ($TP_c$); otherwise, it is a false-positive detection ($FP_c$).
With this, the precision $P_c(k)$ is defined as
\begin{equation}
	P_c(k) = \frac{TP_c(k)}{TP_c(k) + FP_c(k)} \text{,}
\end{equation}
with the index $\text{c}$ denoting the class evaluated.
Recall $R_c(k)$ can be calculated with the false-negative detections $FN_c$ as
\begin{equation}
	R_c(k) = \frac{TP_c(k)}{TP_c(k) + FN_c(k)} \text{.}
\end{equation}

Based on $P_c(k)$ and $R_c(k)$, an average precision $AP_c(k)$ can be determined \cite{b19}.
To also consider the influence of the IoU-threshold, $AP_c(k)$ is again averaged over $K$ IoU thresholds as defined by the detection evaluation metric from Microsoft's Common Object in Context (COCO) \cite{b19}:
\begin{equation}
	AP_c = \sum_{k=1}^{K}\frac{AP_c(k)}{K} \text{.}
\end{equation}

To generate one metric for all classes, the mean average precision $mAP$ calculates the mean over $AP_c$ of all $N$ classes:
\begin{equation}
	mAP = \sum_{c=1}^{N}\frac{AP_c}{N} \text{.}
\end{equation}

The metrics of our UAV dataset are presented in Table~\ref{tab:metrics_panseg}.
Given the large area observed from the aerial perspective, cars, bikes and pedestrians appear rather small in the image.
Consequently, slight detection deviations have a substantial influence on the IoU.
However, the focus of the proposed method is only on detecting objects and not on achieving a perfect IoU.
This is because the labeled RoI has to be larger than the actual object to ensure coverage owing to the limited synchronization precision in the mapping process.
Consequently, the APs reached are sufficient to proof the feasibility of this approach.
Of course, for large-scale applications, these metrics can be improved significantly by increasing the size of the training dataset or by decreasing the IoU threshold.

\section{Sensor Fusion of UAV Camera and Radar}

\subsection{Temporal and Spatial Synchronization}

To map the segmented camera image to the radar RA image in the global frame, both systems must be spatiotemporally synchronized.
Spatial synchronization is achieved using the GNSS data of the radar vehicle and the flight controller of the UAV.
The UTC time from the GNSS receivers can also be used for a precise time synchronization of both systems.

After this step, each camera pixel can be assigned to a region in the RA image.
This sensor fusion is depicted in green in Fig.~\ref{fig:process_chain}.
If the mapping result is not satisfactory due to offsets or positioning inaccuracies, a manual or automated correction can be performed using known reference points in the world plane and the aerial image.

\begin{figure}[tb]
    \centering
	\input{tikz_plots/cart_rois/cart_rois.tex}	
    \vspace{-0.2cm}
    \caption{Section of the radar image transformed to the global frame with the radar in the coordinate origin and the color axis in dB. The RoIs of the detected pedestrian are red.}
    \label{fig:cart_rois}
\end{figure}
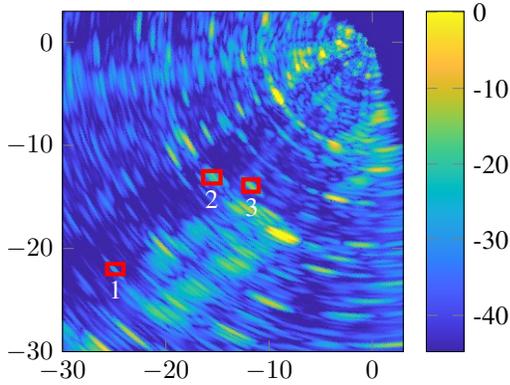

\begin{figure}
    \centering
    \begin{subfigure}{8.8cm}
		\begin{minipage}[l]{0.1cm}
			\caption{}
		\end{minipage}
		\hfill
		\begin{minipage}[r]{8.1cm}
            \hspace{-0.3cm}
			\input{tikz_plots/rd_roi/rd_roi1.tex}
		\end{minipage}
	\end{subfigure}
     \begin{subfigure}{8.8cm}
		\begin{minipage}[l]{0.1cm}
			\caption{}
		\end{minipage}
		\hfill
		\begin{minipage}[r]{8.1cm}
            \hspace{-0.3cm}
			\input{tikz_plots/rd_roi/rd_roi2.tex}
		\end{minipage}
	\end{subfigure}

    \begin{subfigure}{8.8cm}
		\begin{minipage}[l]{0.1cm}
			\caption{}
		\end{minipage}
		\hfill
		\begin{minipage}[r]{8.1cm}
            \hspace{-0.3cm}
			\input{tikz_plots/rd_roi/rd_roi3.tex}
		\end{minipage}
	\end{subfigure}
    \vspace{-0.2cm}
\caption{RD maps with the color axis in dB and the range section of the RoIs in red. a) shows pedestrian 1 in the azimuth section from \qtyrange{-5}{-2}{\degree}, b) shows pedestrian 2 from \qtyrange{-1}{7}{\degree}, and c) shows pedestrian 3 from \qtyrange{-1}{7}{\degree}.}
\label{fig:rd_roi}
\end{figure}
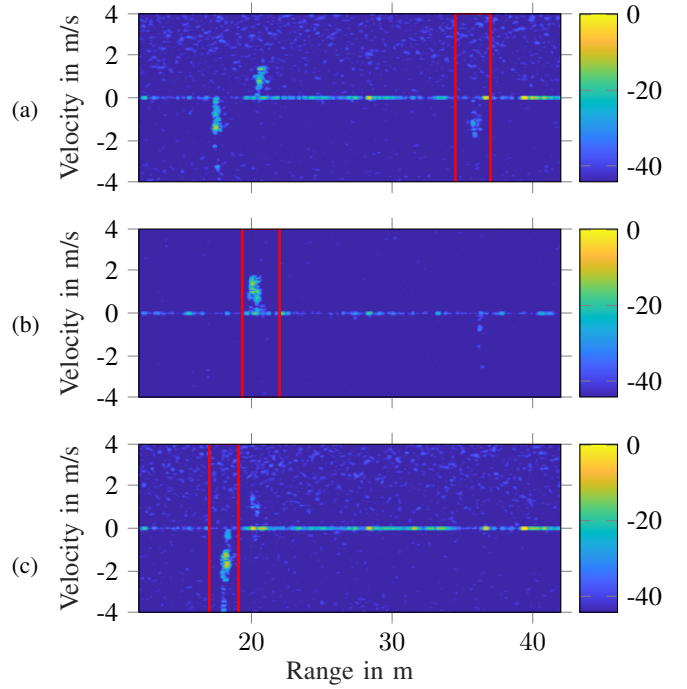
\subsection{RoI Extraction Using Bounding Boxes}

Based on the preceding spatial overlay in the world frame, the panoptic segmentation of the camera image can be directly transferred to the RA image as labels.
This can be realized using a binary mask from the segmented images for each class instance.
Consequently, only instances of the respective class remain labeled, whereas irrelevant parts are removed.

The resulting RoIs within the radar image can be visualized and extracted using rectangular bounding boxes, which allow a larger error margin for imperfect synchronizations.
This is depicted in Fig.~\ref{fig:cart_rois} for the three pedestrians in the radar FOV that are also visible in Fig.~\ref{fig:img_chain}.
For better visualization, the FoV of the depicted radar image is limited in range.
Then, the coordinates of the bounding boxes in the RA image are used to determine the relevant section of the entire radar cube to also consider the object's Doppler information.
The example in Fig.~\ref{fig:cart_rois} also demonstrates that the pedestrians cannot be annotated manually without reference, because they cannot be distinguished from other objects in the RA image.

\section{Generation of Labeled Radar Datasets}

\subsection{Input Data Formats for Neural Networks}

The proposed approach allows the generation of labels for all radar data formats along the processing chain depicted in blue and red in Fig.~\ref{fig:process_chain}, thus providing large flexibility for radar applications.
These range from object detection and classification tasks to radar map segmentation and even ghost target and clutter suppression.
For these tasks, possible input data are shown in Fig.~\ref{fig:process_chain} and include
\begin{itemize}
    \item RD maps, which only require the range information from the extracted RoI for the annotation;
    \item 3D RDA cubes sections that use the extracted and labeled RoI with the respective Doppler content;
    \item Target lists including point clouds derived from the radar cubes using a constant false alarm rate (CFAR) algorithm;
    \item Feature lists with standard target information combined with features cropped from radar cube sections, which are successfully applied in \cite{b20}.
\end{itemize}

Fig.~\ref{fig:rd_roi} demonstrates the effectiveness of the presented approach.
The red rectangle in the RD maps corresponds to the range expansion of the three bounding boxes in Fig.~\ref{fig:cart_rois}.
Within this range expansion, only the azimuth section corresponding to the size of the bounding box is extracted from the radar cube.
With this, a cropped RDA cube containing only the pedestrian can be created.
This small cube, or any of the abovementioned derived data formats, can subsequently be used as training data for a suitable NN architecture.
For visualization in Fig.~\ref{fig:rd_roi}, the cropped 3D RDA cubes are compressed using the maximum value in azimuth dimension.

\subsection{Scalability}

To create large and diverse labeled radar datasets, a trustworthy camera image segmentation is a prerequisite.
For this, an initial and extensive training of the NN for the  camera is required.
Once this network is trained, it can be retrained comparatively easily as needed, resulting in a low long-term overhead and high adaptability.

In our measurement campaign, we recorded \num{248} radar frames in \SI{122}{\s} that contained \num{719} moving pedestrians, from which \num{640} were detected by our panoptic segmentation with an IoU threshold of \SI{50}{\percent}.
Of those, \num{589} were correctly mapped to the RA image to subsequently be extracted as labeled RDA cube section (cf. Fig.~\ref{fig:rd_roi}).

Because we generated \num{589} labeled RDA cube sections of moving pedestrians within \SI{2}{\minute}, this process is only limited by the time required for the recording of measurement data.
No human interaction is required for the complete data processing, as everything can be performed automatically.

Even though the setup presented in this paper is purely static, this approach can also be applied while driving.
For this, a very precise spatiotemporal synchronization with good position estimation of both radar vehicle and UAV is required.

In \cite{b21} a model predictive control (MPC) for object tracking for UAVs is demonstrated.
In this method, the FoV of the UAV's camera remains fixed on a given object like the driving radar vehicle.
With this approach, the UAV can automatically follow the vehicle with a camera FoV suitable for label generation, thus enhancing the creation of large and diverse datasets of dynamic automotive scenarios.

\section{Conclusion}

In this work, we propose an approach for an automated radar data annotation using panoptically segmented aerial images.
For this, the aerial camera images are mapped to the radar's RA images that are transformed to a global coordinate system.
With this approach, the camera image covers a larger section of the radar's FoV.
It also suffers less from occlusion due to the advantageous position, thus enabling a more comprehensive annotation.
The workflow using a purpose-trained UAV camera dataset is presented and discussed in detail.
Additionally, the effectiveness and flexibility of the approach are demonstrated in measurements.
Overall, this approach can be very beneficial for a wide range of environments and radar applications and significantly extends already existing approaches.

In future work, we will extend our approach by the MPC concept from \cite{b21} to create datasets in dynamic traffic scenarios.
The setup can likewise be extended to multiple UAVs to further increase the perspective gain.
The projection approach for aerial images can also be enhanced using geo-referencing to obtain more precise homography results and thus improve the label generation for uneven surfaces. 

\section*{Acknowledgment}
The authors would like to thank the Symeo team from indie Semiconductor for their support with the radar system.

\bibliographystyle{IEEEtran}

\end{document}

%% file: tikz_plots/cart_rois/cart_rois.tex
%
%
\definecolor{mycolor1}{rgb}{0.81815,0.81755,0.72244}%
\definecolor{mycolor2}{rgb}{0.14987,0.65961,0.51859}%
\definecolor{mycolor3}{rgb}{0.97297,0.64899,0.80033}%
\definecolor{mycolor4}{rgb}{0.45380,0.43239,0.82531}%
\definecolor{mycolor5}{rgb}{0.08347,0.13317,0.17339}%
\definecolor{mycolor6}{rgb}{0.39094,0.83138,0.80336}%
\definecolor{mycolor10}{rgb}{1,0,0}%
\begin{tikzpicture}

\begin{axis}[%
scale=0.5,
width=3.566in,
height=3.566in,
at={(0.879in,0.481in)},
scale only axis,
point meta min=-30,
point meta max=15,
axis on top,
xmin=-30,
xmax=3,
ymin=-30,
ymax=3,
axis background/.style={fill=white},
colormap={mymap}{[1pt] rgb(0pt)=(0.2422,0.1504,0.6603); rgb(1pt)=(0.2444,0.1534,0.6728); rgb(2pt)=(0.2464,0.1569,0.6847); rgb(3pt)=(0.2484,0.1607,0.6961); rgb(4pt)=(0.2503,0.1648,0.7071); rgb(5pt)=(0.2522,0.1689,0.7179); rgb(6pt)=(0.254,0.1732,0.7286); rgb(7pt)=(0.2558,0.1773,0.7393); rgb(8pt)=(0.2576,0.1814,0.7501); rgb(9pt)=(0.2594,0.1854,0.761); rgb(11pt)=(0.2628,0.1932,0.7828); rgb(12pt)=(0.2645,0.1972,0.7937); rgb(13pt)=(0.2661,0.2011,0.8043); rgb(14pt)=(0.2676,0.2052,0.8148); rgb(15pt)=(0.2691,0.2094,0.8249); rgb(16pt)=(0.2704,0.2138,0.8346); rgb(17pt)=(0.2717,0.2184,0.8439); rgb(18pt)=(0.2729,0.2231,0.8528); rgb(19pt)=(0.274,0.228,0.8612); rgb(20pt)=(0.2749,0.233,0.8692); rgb(21pt)=(0.2758,0.2382,0.8767); rgb(22pt)=(0.2766,0.2435,0.884); rgb(23pt)=(0.2774,0.2489,0.8908); rgb(24pt)=(0.2781,0.2543,0.8973); rgb(25pt)=(0.2788,0.2598,0.9035); rgb(26pt)=(0.2794,0.2653,0.9094); rgb(27pt)=(0.2798,0.2708,0.915); rgb(28pt)=(0.2802,0.2764,0.9204); rgb(29pt)=(0.2806,0.2819,0.9255); rgb(30pt)=(0.2809,0.2875,0.9305); rgb(31pt)=(0.2811,0.293,0.9352); rgb(32pt)=(0.2813,0.2985,0.9397); rgb(33pt)=(0.2814,0.304,0.9441); rgb(34pt)=(0.2814,0.3095,0.9483); rgb(35pt)=(0.2813,0.315,0.9524); rgb(36pt)=(0.2811,0.3204,0.9563); rgb(37pt)=(0.2809,0.3259,0.96); rgb(38pt)=(0.2807,0.3313,0.9636); rgb(39pt)=(0.2803,0.3367,0.967); rgb(40pt)=(0.2798,0.3421,0.9702); rgb(41pt)=(0.2791,0.3475,0.9733); rgb(42pt)=(0.2784,0.3529,0.9763); rgb(43pt)=(0.2776,0.3583,0.9791); rgb(44pt)=(0.2766,0.3638,0.9817); rgb(45pt)=(0.2754,0.3693,0.984); rgb(46pt)=(0.2741,0.3748,0.9862); rgb(47pt)=(0.2726,0.3804,0.9881); rgb(48pt)=(0.271,0.386,0.9898); rgb(49pt)=(0.2691,0.3916,0.9912); rgb(50pt)=(0.267,0.3973,0.9924); rgb(51pt)=(0.2647,0.403,0.9935); rgb(52pt)=(0.2621,0.4088,0.9946); rgb(53pt)=(0.2591,0.4145,0.9955); rgb(54pt)=(0.2556,0.4203,0.9965); rgb(55pt)=(0.2517,0.4261,0.9974); rgb(56pt)=(0.2473,0.4319,0.9983); rgb(57pt)=(0.2424,0.4378,0.9991); rgb(58pt)=(0.2369,0.4437,0.9996); rgb(59pt)=(0.2311,0.4497,0.9995); rgb(60pt)=(0.225,0.4559,0.9985); rgb(61pt)=(0.2189,0.462,0.9968); rgb(62pt)=(0.2128,0.4682,0.9948); rgb(63pt)=(0.2066,0.4743,0.9926); rgb(64pt)=(0.2006,0.4803,0.9906); rgb(65pt)=(0.195,0.4861,0.9887); rgb(66pt)=(0.1903,0.4919,0.9867); rgb(67pt)=(0.1869,0.4975,0.9844); rgb(68pt)=(0.1847,0.503,0.9819); rgb(69pt)=(0.1831,0.5084,0.9793); rgb(70pt)=(0.1818,0.5138,0.9766); rgb(71pt)=(0.1806,0.5191,0.9738); rgb(72pt)=(0.1795,0.5244,0.9709); rgb(73pt)=(0.1785,0.5296,0.9677); rgb(74pt)=(0.1778,0.5349,0.9641); rgb(75pt)=(0.1773,0.5401,0.9602); rgb(76pt)=(0.1768,0.5452,0.956); rgb(77pt)=(0.1764,0.5504,0.9516); rgb(78pt)=(0.1755,0.5554,0.9473); rgb(79pt)=(0.174,0.5605,0.9432); rgb(80pt)=(0.1716,0.5655,0.9393); rgb(81pt)=(0.1686,0.5705,0.9357); rgb(82pt)=(0.1649,0.5755,0.9323); rgb(83pt)=(0.161,0.5805,0.9289); rgb(84pt)=(0.1573,0.5854,0.9254); rgb(85pt)=(0.154,0.5902,0.9218); rgb(86pt)=(0.1513,0.595,0.9182); rgb(87pt)=(0.1492,0.5997,0.9147); rgb(88pt)=(0.1475,0.6043,0.9113); rgb(89pt)=(0.1461,0.6089,0.908); rgb(90pt)=(0.1446,0.6135,0.905); rgb(91pt)=(0.1429,0.618,0.9022); rgb(92pt)=(0.1408,0.6226,0.8998); rgb(93pt)=(0.1383,0.6272,0.8975); rgb(94pt)=(0.1354,0.6317,0.8953); rgb(95pt)=(0.1321,0.6363,0.8932); rgb(96pt)=(0.1288,0.6408,0.891); rgb(97pt)=(0.1253,0.6453,0.8887); rgb(98pt)=(0.1219,0.6497,0.8862); rgb(99pt)=(0.1185,0.6541,0.8834); rgb(100pt)=(0.1152,0.6584,0.8804); rgb(101pt)=(0.1119,0.6627,0.877); rgb(102pt)=(0.1085,0.6669,0.8734); rgb(103pt)=(0.1048,0.671,0.8695); rgb(104pt)=(0.1009,0.675,0.8653); rgb(105pt)=(0.0964,0.6789,0.8609); rgb(106pt)=(0.0914,0.6828,0.8562); rgb(107pt)=(0.0855,0.6865,0.8513); rgb(108pt)=(0.0789,0.6902,0.8462); rgb(109pt)=(0.0713,0.6938,0.8409); rgb(110pt)=(0.0628,0.6972,0.8355); rgb(111pt)=(0.0535,0.7006,0.8299); rgb(112pt)=(0.0433,0.7039,0.8242); rgb(113pt)=(0.0328,0.7071,0.8183); rgb(114pt)=(0.0234,0.7103,0.8124); rgb(115pt)=(0.0155,0.7133,0.8064); rgb(116pt)=(0.0091,0.7163,0.8003); rgb(117pt)=(0.0046,0.7192,0.7941); rgb(118pt)=(0.0019,0.722,0.7878); rgb(119pt)=(0.0009,0.7248,0.7815); rgb(120pt)=(0.0018,0.7275,0.7752); rgb(121pt)=(0.0046,0.7301,0.7688); rgb(122pt)=(0.0094,0.7327,0.7623); rgb(123pt)=(0.0162,0.7352,0.7558); rgb(124pt)=(0.0253,0.7376,0.7492); rgb(125pt)=(0.0369,0.74,0.7426); rgb(126pt)=(0.0504,0.7423,0.7359); rgb(127pt)=(0.0638,0.7446,0.7292); rgb(128pt)=(0.077,0.7468,0.7224); rgb(129pt)=(0.0899,0.7489,0.7156); rgb(130pt)=(0.1023,0.751,0.7088); rgb(131pt)=(0.1141,0.7531,0.7019); rgb(132pt)=(0.1252,0.7552,0.695); rgb(133pt)=(0.1354,0.7572,0.6881); rgb(134pt)=(0.1448,0.7593,0.6812); rgb(135pt)=(0.1532,0.7614,0.6741); rgb(136pt)=(0.1609,0.7635,0.6671); rgb(137pt)=(0.1678,0.7656,0.6599); rgb(138pt)=(0.1741,0.7678,0.6527); rgb(139pt)=(0.1799,0.7699,0.6454); rgb(140pt)=(0.1853,0.7721,0.6379); rgb(141pt)=(0.1905,0.7743,0.6303); rgb(142pt)=(0.1954,0.7765,0.6225); rgb(143pt)=(0.2003,0.7787,0.6146); rgb(144pt)=(0.2061,0.7808,0.6065); rgb(145pt)=(0.2118,0.7828,0.5983); rgb(146pt)=(0.2178,0.7849,0.5899); rgb(147pt)=(0.2244,0.7869,0.5813); rgb(148pt)=(0.2318,0.7887,0.5725); rgb(149pt)=(0.2401,0.7905,0.5636); rgb(150pt)=(0.2491,0.7922,0.5546); rgb(151pt)=(0.2589,0.7937,0.5454); rgb(152pt)=(0.2695,0.7951,0.536); rgb(153pt)=(0.2809,0.7964,0.5266); rgb(154pt)=(0.2929,0.7975,0.517); rgb(155pt)=(0.3052,0.7985,0.5074); rgb(156pt)=(0.3176,0.7994,0.4975); rgb(157pt)=(0.3301,0.8002,0.4876); rgb(158pt)=(0.3424,0.8009,0.4774); rgb(159pt)=(0.3548,0.8016,0.4669); rgb(160pt)=(0.3671,0.8021,0.4563); rgb(161pt)=(0.3795,0.8026,0.4454); rgb(162pt)=(0.3921,0.8029,0.4344); rgb(163pt)=(0.405,0.8031,0.4233); rgb(164pt)=(0.4184,0.803,0.4122); rgb(165pt)=(0.4322,0.8028,0.4013); rgb(166pt)=(0.4463,0.8024,0.3904); rgb(167pt)=(0.4608,0.8018,0.3797); rgb(168pt)=(0.4753,0.8011,0.3691); rgb(169pt)=(0.4899,0.8002,0.3586); rgb(170pt)=(0.5044,0.7993,0.348); rgb(171pt)=(0.5187,0.7982,0.3374); rgb(172pt)=(0.5329,0.797,0.3267); rgb(173pt)=(0.547,0.7957,0.3159); rgb(175pt)=(0.5748,0.7929,0.2941); rgb(176pt)=(0.5886,0.7913,0.2833); rgb(177pt)=(0.6024,0.7896,0.2726); rgb(178pt)=(0.6161,0.7878,0.2622); rgb(179pt)=(0.6297,0.7859,0.2521); rgb(180pt)=(0.6433,0.7839,0.2423); rgb(181pt)=(0.6567,0.7818,0.2329); rgb(182pt)=(0.6701,0.7796,0.2239); rgb(183pt)=(0.6833,0.7773,0.2155); rgb(184pt)=(0.6963,0.775,0.2075); rgb(185pt)=(0.7091,0.7727,0.1998); rgb(186pt)=(0.7218,0.7703,0.1924); rgb(187pt)=(0.7344,0.7679,0.1852); rgb(188pt)=(0.7468,0.7654,0.1782); rgb(189pt)=(0.759,0.7629,0.1717); rgb(190pt)=(0.771,0.7604,0.1658); rgb(191pt)=(0.7829,0.7579,0.1608); rgb(192pt)=(0.7945,0.7554,0.157); rgb(193pt)=(0.806,0.7529,0.1546); rgb(194pt)=(0.8172,0.7505,0.1535); rgb(195pt)=(0.8281,0.7481,0.1536); rgb(196pt)=(0.8389,0.7457,0.1546); rgb(197pt)=(0.8495,0.7435,0.1564); rgb(198pt)=(0.86,0.7413,0.1587); rgb(199pt)=(0.8703,0.7392,0.1615); rgb(200pt)=(0.8804,0.7372,0.165); rgb(201pt)=(0.8903,0.7353,0.1695); rgb(202pt)=(0.9,0.7336,0.1749); rgb(203pt)=(0.9093,0.7321,0.1815); rgb(204pt)=(0.9184,0.7308,0.189); rgb(205pt)=(0.9272,0.7298,0.1973); rgb(206pt)=(0.9357,0.729,0.2061); rgb(207pt)=(0.944,0.7285,0.2151); rgb(208pt)=(0.9523,0.7284,0.2237); rgb(209pt)=(0.9606,0.7285,0.2312); rgb(210pt)=(0.9689,0.7292,0.2373); rgb(211pt)=(0.977,0.7304,0.2418); rgb(212pt)=(0.9842,0.733,0.2446); rgb(213pt)=(0.99,0.7365,0.2429); rgb(214pt)=(0.9946,0.7407,0.2394); rgb(215pt)=(0.9966,0.7458,0.2351); rgb(216pt)=(0.9971,0.7513,0.2309); rgb(217pt)=(0.9972,0.7569,0.2267); rgb(218pt)=(0.9971,0.7626,0.2224); rgb(219pt)=(0.9969,0.7683,0.2181); rgb(220pt)=(0.9966,0.774,0.2138); rgb(221pt)=(0.9962,0.7798,0.2095); rgb(222pt)=(0.9957,0.7856,0.2053); rgb(223pt)=(0.9949,0.7915,0.2012); rgb(224pt)=(0.9938,0.7974,0.1974); rgb(225pt)=(0.9923,0.8034,0.1939); rgb(226pt)=(0.9906,0.8095,0.1906); rgb(227pt)=(0.9885,0.8156,0.1875); rgb(228pt)=(0.9861,0.8218,0.1846); rgb(229pt)=(0.9835,0.828,0.1817); rgb(230pt)=(0.9807,0.8342,0.1787); rgb(231pt)=(0.9778,0.8404,0.1757); rgb(232pt)=(0.9748,0.8467,0.1726); rgb(233pt)=(0.972,0.8529,0.1695); rgb(234pt)=(0.9694,0.8591,0.1665); rgb(235pt)=(0.9671,0.8654,0.1636); rgb(236pt)=(0.9651,0.8716,0.1608); rgb(237pt)=(0.9634,0.8778,0.1582); rgb(238pt)=(0.9619,0.884,0.1557); rgb(239pt)=(0.9608,0.8902,0.1532); rgb(240pt)=(0.9601,0.8963,0.1507); rgb(241pt)=(0.9596,0.9023,0.148); rgb(242pt)=(0.9595,0.9084,0.145); rgb(243pt)=(0.9597,0.9143,0.1418); rgb(244pt)=(0.9601,0.9203,0.1382); rgb(245pt)=(0.9608,0.9262,0.1344); rgb(246pt)=(0.9618,0.932,0.1304); rgb(247pt)=(0.9629,0.9379,0.1261); rgb(248pt)=(0.9642,0.9437,0.1216); rgb(249pt)=(0.9657,0.9494,0.1168); rgb(250pt)=(0.9674,0.9552,0.1116); rgb(251pt)=(0.9692,0.9609,0.1061); rgb(252pt)=(0.9711,0.9667,0.1001); rgb(253pt)=(0.973,0.9724,0.0938); rgb(254pt)=(0.9749,0.9782,0.0872); rgb(255pt)=(0.9769,0.9839,0.0805)},
colorbar,
colorbar style={ytick={-25.2,-15.2,-5.2, 4.8,14.9}, yticklabels={-40,-30,-20, -10,0}},
]
\addplot [forget plot] graphics [xmin=-80.05, xmax=80.05, ymin=80.05, ymax=-80.05] {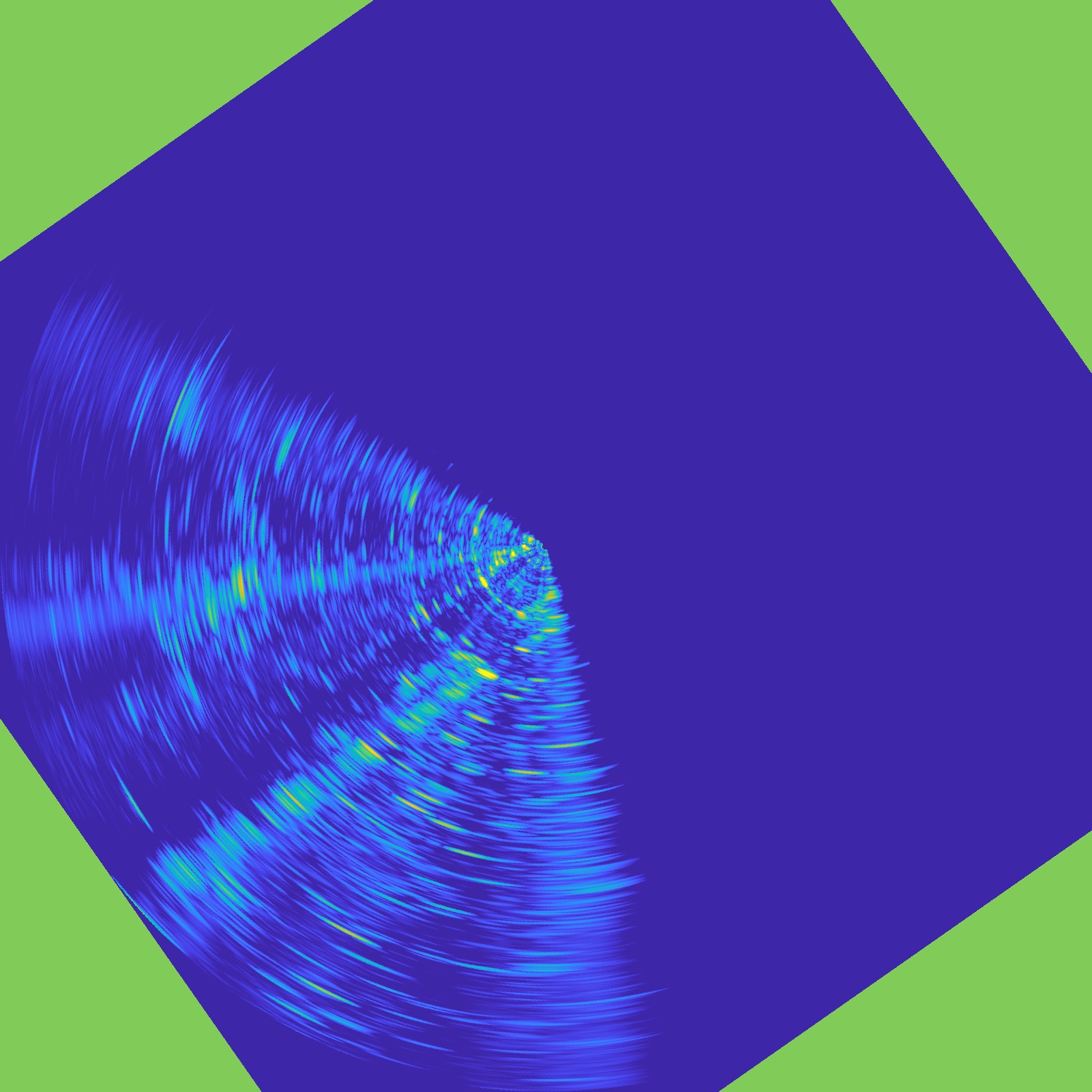};
\draw[line width=1.6pt, draw=mycolor10] (axis cs:-25.7,-22.5) rectangle (axis cs:-24.1,-21.5);
\node[right, align=left, font=\color{white}]
at (axis cs:-26.4,-24) {1};
\draw[line width=1.6pt, draw=mycolor10] (axis cs:-16.4,-13.7) rectangle (axis cs:-14.7,-12.5);
\node[right, align=left, font=\color{white}]
at (axis cs:-17.1,-15.2) {2};
\draw[line width=1.6pt, draw=mycolor10] (axis cs:-12.5,-14.5) rectangle (axis cs:-11.0,-13.3);
\node[right, align=left, font=\color{white}]
at (axis cs:-13.2,-16.0) {3};
\end{axis}
\end{tikzpicture}%

%% file: tikz_plots/rd_roi/rd_roi1.tex
%
%
\begin{tikzpicture}

    \begin{axis}[%
    scale=0.2,
    width=11.04in,
    height=4.4in,
    at={(1.595in,4.769in)},
    scale only axis,
    point meta min=-30,
    point meta max=10,
    axis on top,
    xmin=0.550531089879742,
    xmax=79.9843026296711,
    tick align=outside,
    xlabel style={font=\color{white!15!black}},
    xtick={20, 30, 40},
    xticklabels={},
    ymin=-3.99308131128253,
    ymax=3.93068941579374,
    ylabel style={font=\color{white!15!black}},
    ylabel={Velocity in m/s},
    axis background/.style={fill=white},
    xminorgrids,
    yminorgrids,
    xmin=12,
    xmax=42,
    ymin=-3.9,
    ymax=3.9,
    ytick={-3.9,-2,0,2,3.9},
    yticklabels={-4,-2,0,2,4},
    ylabel style={font=\color{white!15!black}, yshift=-0.34cm},
    ylabel={Velocity in m/s},
    colormap={mymap}{[1pt] rgb(0pt)=(0.2422,0.1504,0.6603); rgb(1pt)=(0.2444,0.1534,0.6728); rgb(2pt)=(0.2464,0.1569,0.6847); rgb(3pt)=(0.2484,0.1607,0.6961); rgb(4pt)=(0.2503,0.1648,0.7071); rgb(5pt)=(0.2522,0.1689,0.7179); rgb(6pt)=(0.254,0.1732,0.7286); rgb(7pt)=(0.2558,0.1773,0.7393); rgb(8pt)=(0.2576,0.1814,0.7501); rgb(9pt)=(0.2594,0.1854,0.761); rgb(11pt)=(0.2628,0.1932,0.7828); rgb(12pt)=(0.2645,0.1972,0.7937); rgb(13pt)=(0.2661,0.2011,0.8043); rgb(14pt)=(0.2676,0.2052,0.8148); rgb(15pt)=(0.2691,0.2094,0.8249); rgb(16pt)=(0.2704,0.2138,0.8346); rgb(17pt)=(0.2717,0.2184,0.8439); rgb(18pt)=(0.2729,0.2231,0.8528); rgb(19pt)=(0.274,0.228,0.8612); rgb(20pt)=(0.2749,0.233,0.8692); rgb(21pt)=(0.2758,0.2382,0.8767); rgb(22pt)=(0.2766,0.2435,0.884); rgb(23pt)=(0.2774,0.2489,0.8908); rgb(24pt)=(0.2781,0.2543,0.8973); rgb(25pt)=(0.2788,0.2598,0.9035); rgb(26pt)=(0.2794,0.2653,0.9094); rgb(27pt)=(0.2798,0.2708,0.915); rgb(28pt)=(0.2802,0.2764,0.9204); rgb(29pt)=(0.2806,0.2819,0.9255); rgb(30pt)=(0.2809,0.2875,0.9305); rgb(31pt)=(0.2811,0.293,0.9352); rgb(32pt)=(0.2813,0.2985,0.9397); rgb(33pt)=(0.2814,0.304,0.9441); rgb(34pt)=(0.2814,0.3095,0.9483); rgb(35pt)=(0.2813,0.315,0.9524); rgb(36pt)=(0.2811,0.3204,0.9563); rgb(37pt)=(0.2809,0.3259,0.96); rgb(38pt)=(0.2807,0.3313,0.9636); rgb(39pt)=(0.2803,0.3367,0.967); rgb(40pt)=(0.2798,0.3421,0.9702); rgb(41pt)=(0.2791,0.3475,0.9733); rgb(42pt)=(0.2784,0.3529,0.9763); rgb(43pt)=(0.2776,0.3583,0.9791); rgb(44pt)=(0.2766,0.3638,0.9817); rgb(45pt)=(0.2754,0.3693,0.984); rgb(46pt)=(0.2741,0.3748,0.9862); rgb(47pt)=(0.2726,0.3804,0.9881); rgb(48pt)=(0.271,0.386,0.9898); rgb(49pt)=(0.2691,0.3916,0.9912); rgb(50pt)=(0.267,0.3973,0.9924); rgb(51pt)=(0.2647,0.403,0.9935); rgb(52pt)=(0.2621,0.4088,0.9946); rgb(53pt)=(0.2591,0.4145,0.9955); rgb(54pt)=(0.2556,0.4203,0.9965); rgb(55pt)=(0.2517,0.4261,0.9974); rgb(56pt)=(0.2473,0.4319,0.9983); rgb(57pt)=(0.2424,0.4378,0.9991); rgb(58pt)=(0.2369,0.4437,0.9996); rgb(59pt)=(0.2311,0.4497,0.9995); rgb(60pt)=(0.225,0.4559,0.9985); rgb(61pt)=(0.2189,0.462,0.9968); rgb(62pt)=(0.2128,0.4682,0.9948); rgb(63pt)=(0.2066,0.4743,0.9926); rgb(64pt)=(0.2006,0.4803,0.9906); rgb(65pt)=(0.195,0.4861,0.9887); rgb(66pt)=(0.1903,0.4919,0.9867); rgb(67pt)=(0.1869,0.4975,0.9844); rgb(68pt)=(0.1847,0.503,0.9819); rgb(69pt)=(0.1831,0.5084,0.9793); rgb(70pt)=(0.1818,0.5138,0.9766); rgb(71pt)=(0.1806,0.5191,0.9738); rgb(72pt)=(0.1795,0.5244,0.9709); rgb(73pt)=(0.1785,0.5296,0.9677); rgb(74pt)=(0.1778,0.5349,0.9641); rgb(75pt)=(0.1773,0.5401,0.9602); rgb(76pt)=(0.1768,0.5452,0.956); rgb(77pt)=(0.1764,0.5504,0.9516); rgb(78pt)=(0.1755,0.5554,0.9473); rgb(79pt)=(0.174,0.5605,0.9432); rgb(80pt)=(0.1716,0.5655,0.9393); rgb(81pt)=(0.1686,0.5705,0.9357); rgb(82pt)=(0.1649,0.5755,0.9323); rgb(83pt)=(0.161,0.5805,0.9289); rgb(84pt)=(0.1573,0.5854,0.9254); rgb(85pt)=(0.154,0.5902,0.9218); rgb(86pt)=(0.1513,0.595,0.9182); rgb(87pt)=(0.1492,0.5997,0.9147); rgb(88pt)=(0.1475,0.6043,0.9113); rgb(89pt)=(0.1461,0.6089,0.908); rgb(90pt)=(0.1446,0.6135,0.905); rgb(91pt)=(0.1429,0.618,0.9022); rgb(92pt)=(0.1408,0.6226,0.8998); rgb(93pt)=(0.1383,0.6272,0.8975); rgb(94pt)=(0.1354,0.6317,0.8953); rgb(95pt)=(0.1321,0.6363,0.8932); rgb(96pt)=(0.1288,0.6408,0.891); rgb(97pt)=(0.1253,0.6453,0.8887); rgb(98pt)=(0.1219,0.6497,0.8862); rgb(99pt)=(0.1185,0.6541,0.8834); rgb(100pt)=(0.1152,0.6584,0.8804); rgb(101pt)=(0.1119,0.6627,0.877); rgb(102pt)=(0.1085,0.6669,0.8734); rgb(103pt)=(0.1048,0.671,0.8695); rgb(104pt)=(0.1009,0.675,0.8653); rgb(105pt)=(0.0964,0.6789,0.8609); rgb(106pt)=(0.0914,0.6828,0.8562); rgb(107pt)=(0.0855,0.6865,0.8513); rgb(108pt)=(0.0789,0.6902,0.8462); rgb(109pt)=(0.0713,0.6938,0.8409); rgb(110pt)=(0.0628,0.6972,0.8355); rgb(111pt)=(0.0535,0.7006,0.8299); rgb(112pt)=(0.0433,0.7039,0.8242); rgb(113pt)=(0.0328,0.7071,0.8183); rgb(114pt)=(0.0234,0.7103,0.8124); rgb(115pt)=(0.0155,0.7133,0.8064); rgb(116pt)=(0.0091,0.7163,0.8003); rgb(117pt)=(0.0046,0.7192,0.7941); rgb(118pt)=(0.0019,0.722,0.7878); rgb(119pt)=(0.0009,0.7248,0.7815); rgb(120pt)=(0.0018,0.7275,0.7752); rgb(121pt)=(0.0046,0.7301,0.7688); rgb(122pt)=(0.0094,0.7327,0.7623); rgb(123pt)=(0.0162,0.7352,0.7558); rgb(124pt)=(0.0253,0.7376,0.7492); rgb(125pt)=(0.0369,0.74,0.7426); rgb(126pt)=(0.0504,0.7423,0.7359); rgb(127pt)=(0.0638,0.7446,0.7292); rgb(128pt)=(0.077,0.7468,0.7224); rgb(129pt)=(0.0899,0.7489,0.7156); rgb(130pt)=(0.1023,0.751,0.7088); rgb(131pt)=(0.1141,0.7531,0.7019); rgb(132pt)=(0.1252,0.7552,0.695); rgb(133pt)=(0.1354,0.7572,0.6881); rgb(134pt)=(0.1448,0.7593,0.6812); rgb(135pt)=(0.1532,0.7614,0.6741); rgb(136pt)=(0.1609,0.7635,0.6671); rgb(137pt)=(0.1678,0.7656,0.6599); rgb(138pt)=(0.1741,0.7678,0.6527); rgb(139pt)=(0.1799,0.7699,0.6454); rgb(140pt)=(0.1853,0.7721,0.6379); rgb(141pt)=(0.1905,0.7743,0.6303); rgb(142pt)=(0.1954,0.7765,0.6225); rgb(143pt)=(0.2003,0.7787,0.6146); rgb(144pt)=(0.2061,0.7808,0.6065); rgb(145pt)=(0.2118,0.7828,0.5983); rgb(146pt)=(0.2178,0.7849,0.5899); rgb(147pt)=(0.2244,0.7869,0.5813); rgb(148pt)=(0.2318,0.7887,0.5725); rgb(149pt)=(0.2401,0.7905,0.5636); rgb(150pt)=(0.2491,0.7922,0.5546); rgb(151pt)=(0.2589,0.7937,0.5454); rgb(152pt)=(0.2695,0.7951,0.536); rgb(153pt)=(0.2809,0.7964,0.5266); rgb(154pt)=(0.2929,0.7975,0.517); rgb(155pt)=(0.3052,0.7985,0.5074); rgb(156pt)=(0.3176,0.7994,0.4975); rgb(157pt)=(0.3301,0.8002,0.4876); rgb(158pt)=(0.3424,0.8009,0.4774); rgb(159pt)=(0.3548,0.8016,0.4669); rgb(160pt)=(0.3671,0.8021,0.4563); rgb(161pt)=(0.3795,0.8026,0.4454); rgb(162pt)=(0.3921,0.8029,0.4344); rgb(163pt)=(0.405,0.8031,0.4233); rgb(164pt)=(0.4184,0.803,0.4122); rgb(165pt)=(0.4322,0.8028,0.4013); rgb(166pt)=(0.4463,0.8024,0.3904); rgb(167pt)=(0.4608,0.8018,0.3797); rgb(168pt)=(0.4753,0.8011,0.3691); rgb(169pt)=(0.4899,0.8002,0.3586); rgb(170pt)=(0.5044,0.7993,0.348); rgb(171pt)=(0.5187,0.7982,0.3374); rgb(172pt)=(0.5329,0.797,0.3267); rgb(173pt)=(0.547,0.7957,0.3159); rgb(175pt)=(0.5748,0.7929,0.2941); rgb(176pt)=(0.5886,0.7913,0.2833); rgb(177pt)=(0.6024,0.7896,0.2726); rgb(178pt)=(0.6161,0.7878,0.2622); rgb(179pt)=(0.6297,0.7859,0.2521); rgb(180pt)=(0.6433,0.7839,0.2423); rgb(181pt)=(0.6567,0.7818,0.2329); rgb(182pt)=(0.6701,0.7796,0.2239); rgb(183pt)=(0.6833,0.7773,0.2155); rgb(184pt)=(0.6963,0.775,0.2075); rgb(185pt)=(0.7091,0.7727,0.1998); rgb(186pt)=(0.7218,0.7703,0.1924); rgb(187pt)=(0.7344,0.7679,0.1852); rgb(188pt)=(0.7468,0.7654,0.1782); rgb(189pt)=(0.759,0.7629,0.1717); rgb(190pt)=(0.771,0.7604,0.1658); rgb(191pt)=(0.7829,0.7579,0.1608); rgb(192pt)=(0.7945,0.7554,0.157); rgb(193pt)=(0.806,0.7529,0.1546); rgb(194pt)=(0.8172,0.7505,0.1535); rgb(195pt)=(0.8281,0.7481,0.1536); rgb(196pt)=(0.8389,0.7457,0.1546); rgb(197pt)=(0.8495,0.7435,0.1564); rgb(198pt)=(0.86,0.7413,0.1587); rgb(199pt)=(0.8703,0.7392,0.1615); rgb(200pt)=(0.8804,0.7372,0.165); rgb(201pt)=(0.8903,0.7353,0.1695); rgb(202pt)=(0.9,0.7336,0.1749); rgb(203pt)=(0.9093,0.7321,0.1815); rgb(204pt)=(0.9184,0.7308,0.189); rgb(205pt)=(0.9272,0.7298,0.1973); rgb(206pt)=(0.9357,0.729,0.2061); rgb(207pt)=(0.944,0.7285,0.2151); rgb(208pt)=(0.9523,0.7284,0.2237); rgb(209pt)=(0.9606,0.7285,0.2312); rgb(210pt)=(0.9689,0.7292,0.2373); rgb(211pt)=(0.977,0.7304,0.2418); rgb(212pt)=(0.9842,0.733,0.2446); rgb(213pt)=(0.99,0.7365,0.2429); rgb(214pt)=(0.9946,0.7407,0.2394); rgb(215pt)=(0.9966,0.7458,0.2351); rgb(216pt)=(0.9971,0.7513,0.2309); rgb(217pt)=(0.9972,0.7569,0.2267); rgb(218pt)=(0.9971,0.7626,0.2224); rgb(219pt)=(0.9969,0.7683,0.2181); rgb(220pt)=(0.9966,0.774,0.2138); rgb(221pt)=(0.9962,0.7798,0.2095); rgb(222pt)=(0.9957,0.7856,0.2053); rgb(223pt)=(0.9949,0.7915,0.2012); rgb(224pt)=(0.9938,0.7974,0.1974); rgb(225pt)=(0.9923,0.8034,0.1939); rgb(226pt)=(0.9906,0.8095,0.1906); rgb(227pt)=(0.9885,0.8156,0.1875); rgb(228pt)=(0.9861,0.8218,0.1846); rgb(229pt)=(0.9835,0.828,0.1817); rgb(230pt)=(0.9807,0.8342,0.1787); rgb(231pt)=(0.9778,0.8404,0.1757); rgb(232pt)=(0.9748,0.8467,0.1726); rgb(233pt)=(0.972,0.8529,0.1695); rgb(234pt)=(0.9694,0.8591,0.1665); rgb(235pt)=(0.9671,0.8654,0.1636); rgb(236pt)=(0.9651,0.8716,0.1608); rgb(237pt)=(0.9634,0.8778,0.1582); rgb(238pt)=(0.9619,0.884,0.1557); rgb(239pt)=(0.9608,0.8902,0.1532); rgb(240pt)=(0.9601,0.8963,0.1507); rgb(241pt)=(0.9596,0.9023,0.148); rgb(242pt)=(0.9595,0.9084,0.145); rgb(243pt)=(0.9597,0.9143,0.1418); rgb(244pt)=(0.9601,0.9203,0.1382); rgb(245pt)=(0.9608,0.9262,0.1344); rgb(246pt)=(0.9618,0.932,0.1304); rgb(247pt)=(0.9629,0.9379,0.1261); rgb(248pt)=(0.9642,0.9437,0.1216); rgb(249pt)=(0.9657,0.9494,0.1168); rgb(250pt)=(0.9674,0.9552,0.1116); rgb(251pt)=(0.9692,0.9609,0.1061); rgb(252pt)=(0.9711,0.9667,0.1001); rgb(253pt)=(0.973,0.9724,0.0938); rgb(254pt)=(0.9749,0.9782,0.0872); rgb(255pt)=(0.9769,0.9839,0.0805)},
    colorbar,
    colorbar style={ytick={-26.2,-8.2,9.8}, yticklabels={-40,-20,0}, style={xshift=-0.2cm}},
    ]
    
    \addplot [forget plot] graphics [xmin=0.550531089879742, xmax=79.9843026296711, ymin=-3.99308131128253, ymax=3.93068941579374] {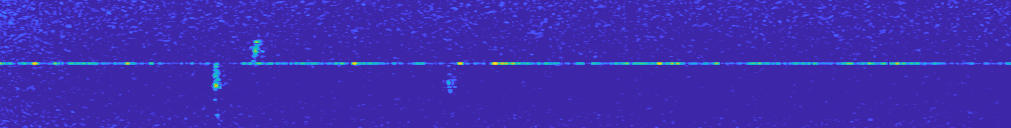};
    \draw[line width=1.0pt, draw=red] (axis cs:34.5,-3.99308131128253) rectangle (axis cs:37,3.93068941579374);
    \end{axis}
    
    \end{tikzpicture}%

%% file: tikz_plots/rd_roi/rd_roi2.tex
%
%
\begin{tikzpicture}

\begin{axis}[%
scale=0.2,
width=11.04in,
height=4.4in,
at={(1.595in,4.769in)},
scale only axis,
point meta min=-30,
point meta max=10,
axis on top,
xmin=0.550531089879742,
xmax=79.9843026296711,
tick align=outside,
xtick={20, 30, 40},
xticklabels={},
ymin=-3.99308131128253,
ymax=3.93068941579374,
ylabel style={font=\color{white!15!black}},
ylabel={Velocity in m/s},
axis background/.style={fill=white},
xminorgrids,
yminorgrids,
xmin=12,
xmax=42,
ymin=-3.9,
ymax=3.9,
ytick={-3.9,-2,0,2,3.9},
yticklabels={-4,-2,0,2,4},
ylabel style={font=\color{white!15!black}, yshift=-0.34cm},
ylabel={Velocity in m/s},
colormap={mymap}{[1pt] rgb(0pt)=(0.2422,0.1504,0.6603); rgb(1pt)=(0.2444,0.1534,0.6728); rgb(2pt)=(0.2464,0.1569,0.6847); rgb(3pt)=(0.2484,0.1607,0.6961); rgb(4pt)=(0.2503,0.1648,0.7071); rgb(5pt)=(0.2522,0.1689,0.7179); rgb(6pt)=(0.254,0.1732,0.7286); rgb(7pt)=(0.2558,0.1773,0.7393); rgb(8pt)=(0.2576,0.1814,0.7501); rgb(9pt)=(0.2594,0.1854,0.761); rgb(11pt)=(0.2628,0.1932,0.7828); rgb(12pt)=(0.2645,0.1972,0.7937); rgb(13pt)=(0.2661,0.2011,0.8043); rgb(14pt)=(0.2676,0.2052,0.8148); rgb(15pt)=(0.2691,0.2094,0.8249); rgb(16pt)=(0.2704,0.2138,0.8346); rgb(17pt)=(0.2717,0.2184,0.8439); rgb(18pt)=(0.2729,0.2231,0.8528); rgb(19pt)=(0.274,0.228,0.8612); rgb(20pt)=(0.2749,0.233,0.8692); rgb(21pt)=(0.2758,0.2382,0.8767); rgb(22pt)=(0.2766,0.2435,0.884); rgb(23pt)=(0.2774,0.2489,0.8908); rgb(24pt)=(0.2781,0.2543,0.8973); rgb(25pt)=(0.2788,0.2598,0.9035); rgb(26pt)=(0.2794,0.2653,0.9094); rgb(27pt)=(0.2798,0.2708,0.915); rgb(28pt)=(0.2802,0.2764,0.9204); rgb(29pt)=(0.2806,0.2819,0.9255); rgb(30pt)=(0.2809,0.2875,0.9305); rgb(31pt)=(0.2811,0.293,0.9352); rgb(32pt)=(0.2813,0.2985,0.9397); rgb(33pt)=(0.2814,0.304,0.9441); rgb(34pt)=(0.2814,0.3095,0.9483); rgb(35pt)=(0.2813,0.315,0.9524); rgb(36pt)=(0.2811,0.3204,0.9563); rgb(37pt)=(0.2809,0.3259,0.96); rgb(38pt)=(0.2807,0.3313,0.9636); rgb(39pt)=(0.2803,0.3367,0.967); rgb(40pt)=(0.2798,0.3421,0.9702); rgb(41pt)=(0.2791,0.3475,0.9733); rgb(42pt)=(0.2784,0.3529,0.9763); rgb(43pt)=(0.2776,0.3583,0.9791); rgb(44pt)=(0.2766,0.3638,0.9817); rgb(45pt)=(0.2754,0.3693,0.984); rgb(46pt)=(0.2741,0.3748,0.9862); rgb(47pt)=(0.2726,0.3804,0.9881); rgb(48pt)=(0.271,0.386,0.9898); rgb(49pt)=(0.2691,0.3916,0.9912); rgb(50pt)=(0.267,0.3973,0.9924); rgb(51pt)=(0.2647,0.403,0.9935); rgb(52pt)=(0.2621,0.4088,0.9946); rgb(53pt)=(0.2591,0.4145,0.9955); rgb(54pt)=(0.2556,0.4203,0.9965); rgb(55pt)=(0.2517,0.4261,0.9974); rgb(56pt)=(0.2473,0.4319,0.9983); rgb(57pt)=(0.2424,0.4378,0.9991); rgb(58pt)=(0.2369,0.4437,0.9996); rgb(59pt)=(0.2311,0.4497,0.9995); rgb(60pt)=(0.225,0.4559,0.9985); rgb(61pt)=(0.2189,0.462,0.9968); rgb(62pt)=(0.2128,0.4682,0.9948); rgb(63pt)=(0.2066,0.4743,0.9926); rgb(64pt)=(0.2006,0.4803,0.9906); rgb(65pt)=(0.195,0.4861,0.9887); rgb(66pt)=(0.1903,0.4919,0.9867); rgb(67pt)=(0.1869,0.4975,0.9844); rgb(68pt)=(0.1847,0.503,0.9819); rgb(69pt)=(0.1831,0.5084,0.9793); rgb(70pt)=(0.1818,0.5138,0.9766); rgb(71pt)=(0.1806,0.5191,0.9738); rgb(72pt)=(0.1795,0.5244,0.9709); rgb(73pt)=(0.1785,0.5296,0.9677); rgb(74pt)=(0.1778,0.5349,0.9641); rgb(75pt)=(0.1773,0.5401,0.9602); rgb(76pt)=(0.1768,0.5452,0.956); rgb(77pt)=(0.1764,0.5504,0.9516); rgb(78pt)=(0.1755,0.5554,0.9473); rgb(79pt)=(0.174,0.5605,0.9432); rgb(80pt)=(0.1716,0.5655,0.9393); rgb(81pt)=(0.1686,0.5705,0.9357); rgb(82pt)=(0.1649,0.5755,0.9323); rgb(83pt)=(0.161,0.5805,0.9289); rgb(84pt)=(0.1573,0.5854,0.9254); rgb(85pt)=(0.154,0.5902,0.9218); rgb(86pt)=(0.1513,0.595,0.9182); rgb(87pt)=(0.1492,0.5997,0.9147); rgb(88pt)=(0.1475,0.6043,0.9113); rgb(89pt)=(0.1461,0.6089,0.908); rgb(90pt)=(0.1446,0.6135,0.905); rgb(91pt)=(0.1429,0.618,0.9022); rgb(92pt)=(0.1408,0.6226,0.8998); rgb(93pt)=(0.1383,0.6272,0.8975); rgb(94pt)=(0.1354,0.6317,0.8953); rgb(95pt)=(0.1321,0.6363,0.8932); rgb(96pt)=(0.1288,0.6408,0.891); rgb(97pt)=(0.1253,0.6453,0.8887); rgb(98pt)=(0.1219,0.6497,0.8862); rgb(99pt)=(0.1185,0.6541,0.8834); rgb(100pt)=(0.1152,0.6584,0.8804); rgb(101pt)=(0.1119,0.6627,0.877); rgb(102pt)=(0.1085,0.6669,0.8734); rgb(103pt)=(0.1048,0.671,0.8695); rgb(104pt)=(0.1009,0.675,0.8653); rgb(105pt)=(0.0964,0.6789,0.8609); rgb(106pt)=(0.0914,0.6828,0.8562); rgb(107pt)=(0.0855,0.6865,0.8513); rgb(108pt)=(0.0789,0.6902,0.8462); rgb(109pt)=(0.0713,0.6938,0.8409); rgb(110pt)=(0.0628,0.6972,0.8355); rgb(111pt)=(0.0535,0.7006,0.8299); rgb(112pt)=(0.0433,0.7039,0.8242); rgb(113pt)=(0.0328,0.7071,0.8183); rgb(114pt)=(0.0234,0.7103,0.8124); rgb(115pt)=(0.0155,0.7133,0.8064); rgb(116pt)=(0.0091,0.7163,0.8003); rgb(117pt)=(0.0046,0.7192,0.7941); rgb(118pt)=(0.0019,0.722,0.7878); rgb(119pt)=(0.0009,0.7248,0.7815); rgb(120pt)=(0.0018,0.7275,0.7752); rgb(121pt)=(0.0046,0.7301,0.7688); rgb(122pt)=(0.0094,0.7327,0.7623); rgb(123pt)=(0.0162,0.7352,0.7558); rgb(124pt)=(0.0253,0.7376,0.7492); rgb(125pt)=(0.0369,0.74,0.7426); rgb(126pt)=(0.0504,0.7423,0.7359); rgb(127pt)=(0.0638,0.7446,0.7292); rgb(128pt)=(0.077,0.7468,0.7224); rgb(129pt)=(0.0899,0.7489,0.7156); rgb(130pt)=(0.1023,0.751,0.7088); rgb(131pt)=(0.1141,0.7531,0.7019); rgb(132pt)=(0.1252,0.7552,0.695); rgb(133pt)=(0.1354,0.7572,0.6881); rgb(134pt)=(0.1448,0.7593,0.6812); rgb(135pt)=(0.1532,0.7614,0.6741); rgb(136pt)=(0.1609,0.7635,0.6671); rgb(137pt)=(0.1678,0.7656,0.6599); rgb(138pt)=(0.1741,0.7678,0.6527); rgb(139pt)=(0.1799,0.7699,0.6454); rgb(140pt)=(0.1853,0.7721,0.6379); rgb(141pt)=(0.1905,0.7743,0.6303); rgb(142pt)=(0.1954,0.7765,0.6225); rgb(143pt)=(0.2003,0.7787,0.6146); rgb(144pt)=(0.2061,0.7808,0.6065); rgb(145pt)=(0.2118,0.7828,0.5983); rgb(146pt)=(0.2178,0.7849,0.5899); rgb(147pt)=(0.2244,0.7869,0.5813); rgb(148pt)=(0.2318,0.7887,0.5725); rgb(149pt)=(0.2401,0.7905,0.5636); rgb(150pt)=(0.2491,0.7922,0.5546); rgb(151pt)=(0.2589,0.7937,0.5454); rgb(152pt)=(0.2695,0.7951,0.536); rgb(153pt)=(0.2809,0.7964,0.5266); rgb(154pt)=(0.2929,0.7975,0.517); rgb(155pt)=(0.3052,0.7985,0.5074); rgb(156pt)=(0.3176,0.7994,0.4975); rgb(157pt)=(0.3301,0.8002,0.4876); rgb(158pt)=(0.3424,0.8009,0.4774); rgb(159pt)=(0.3548,0.8016,0.4669); rgb(160pt)=(0.3671,0.8021,0.4563); rgb(161pt)=(0.3795,0.8026,0.4454); rgb(162pt)=(0.3921,0.8029,0.4344); rgb(163pt)=(0.405,0.8031,0.4233); rgb(164pt)=(0.4184,0.803,0.4122); rgb(165pt)=(0.4322,0.8028,0.4013); rgb(166pt)=(0.4463,0.8024,0.3904); rgb(167pt)=(0.4608,0.8018,0.3797); rgb(168pt)=(0.4753,0.8011,0.3691); rgb(169pt)=(0.4899,0.8002,0.3586); rgb(170pt)=(0.5044,0.7993,0.348); rgb(171pt)=(0.5187,0.7982,0.3374); rgb(172pt)=(0.5329,0.797,0.3267); rgb(173pt)=(0.547,0.7957,0.3159); rgb(175pt)=(0.5748,0.7929,0.2941); rgb(176pt)=(0.5886,0.7913,0.2833); rgb(177pt)=(0.6024,0.7896,0.2726); rgb(178pt)=(0.6161,0.7878,0.2622); rgb(179pt)=(0.6297,0.7859,0.2521); rgb(180pt)=(0.6433,0.7839,0.2423); rgb(181pt)=(0.6567,0.7818,0.2329); rgb(182pt)=(0.6701,0.7796,0.2239); rgb(183pt)=(0.6833,0.7773,0.2155); rgb(184pt)=(0.6963,0.775,0.2075); rgb(185pt)=(0.7091,0.7727,0.1998); rgb(186pt)=(0.7218,0.7703,0.1924); rgb(187pt)=(0.7344,0.7679,0.1852); rgb(188pt)=(0.7468,0.7654,0.1782); rgb(189pt)=(0.759,0.7629,0.1717); rgb(190pt)=(0.771,0.7604,0.1658); rgb(191pt)=(0.7829,0.7579,0.1608); rgb(192pt)=(0.7945,0.7554,0.157); rgb(193pt)=(0.806,0.7529,0.1546); rgb(194pt)=(0.8172,0.7505,0.1535); rgb(195pt)=(0.8281,0.7481,0.1536); rgb(196pt)=(0.8389,0.7457,0.1546); rgb(197pt)=(0.8495,0.7435,0.1564); rgb(198pt)=(0.86,0.7413,0.1587); rgb(199pt)=(0.8703,0.7392,0.1615); rgb(200pt)=(0.8804,0.7372,0.165); rgb(201pt)=(0.8903,0.7353,0.1695); rgb(202pt)=(0.9,0.7336,0.1749); rgb(203pt)=(0.9093,0.7321,0.1815); rgb(204pt)=(0.9184,0.7308,0.189); rgb(205pt)=(0.9272,0.7298,0.1973); rgb(206pt)=(0.9357,0.729,0.2061); rgb(207pt)=(0.944,0.7285,0.2151); rgb(208pt)=(0.9523,0.7284,0.2237); rgb(209pt)=(0.9606,0.7285,0.2312); rgb(210pt)=(0.9689,0.7292,0.2373); rgb(211pt)=(0.977,0.7304,0.2418); rgb(212pt)=(0.9842,0.733,0.2446); rgb(213pt)=(0.99,0.7365,0.2429); rgb(214pt)=(0.9946,0.7407,0.2394); rgb(215pt)=(0.9966,0.7458,0.2351); rgb(216pt)=(0.9971,0.7513,0.2309); rgb(217pt)=(0.9972,0.7569,0.2267); rgb(218pt)=(0.9971,0.7626,0.2224); rgb(219pt)=(0.9969,0.7683,0.2181); rgb(220pt)=(0.9966,0.774,0.2138); rgb(221pt)=(0.9962,0.7798,0.2095); rgb(222pt)=(0.9957,0.7856,0.2053); rgb(223pt)=(0.9949,0.7915,0.2012); rgb(224pt)=(0.9938,0.7974,0.1974); rgb(225pt)=(0.9923,0.8034,0.1939); rgb(226pt)=(0.9906,0.8095,0.1906); rgb(227pt)=(0.9885,0.8156,0.1875); rgb(228pt)=(0.9861,0.8218,0.1846); rgb(229pt)=(0.9835,0.828,0.1817); rgb(230pt)=(0.9807,0.8342,0.1787); rgb(231pt)=(0.9778,0.8404,0.1757); rgb(232pt)=(0.9748,0.8467,0.1726); rgb(233pt)=(0.972,0.8529,0.1695); rgb(234pt)=(0.9694,0.8591,0.1665); rgb(235pt)=(0.9671,0.8654,0.1636); rgb(236pt)=(0.9651,0.8716,0.1608); rgb(237pt)=(0.9634,0.8778,0.1582); rgb(238pt)=(0.9619,0.884,0.1557); rgb(239pt)=(0.9608,0.8902,0.1532); rgb(240pt)=(0.9601,0.8963,0.1507); rgb(241pt)=(0.9596,0.9023,0.148); rgb(242pt)=(0.9595,0.9084,0.145); rgb(243pt)=(0.9597,0.9143,0.1418); rgb(244pt)=(0.9601,0.9203,0.1382); rgb(245pt)=(0.9608,0.9262,0.1344); rgb(246pt)=(0.9618,0.932,0.1304); rgb(247pt)=(0.9629,0.9379,0.1261); rgb(248pt)=(0.9642,0.9437,0.1216); rgb(249pt)=(0.9657,0.9494,0.1168); rgb(250pt)=(0.9674,0.9552,0.1116); rgb(251pt)=(0.9692,0.9609,0.1061); rgb(252pt)=(0.9711,0.9667,0.1001); rgb(253pt)=(0.973,0.9724,0.0938); rgb(254pt)=(0.9749,0.9782,0.0872); rgb(255pt)=(0.9769,0.9839,0.0805)},
colorbar,
colorbar style={ytick={-26.2,-8.2,9.8}, yticklabels={-40,-20,0}, style={xshift=-0.2cm}},
]

\addplot [forget plot] graphics [xmin=0.550531089879742, xmax=79.9843026296711, ymin=-3.99308131128253, ymax=3.93068941579374] {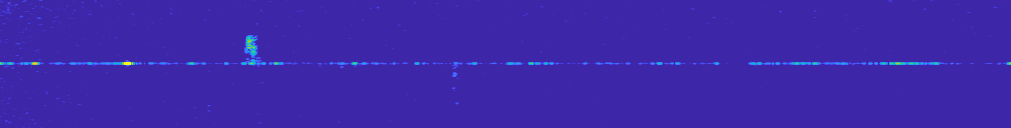};
\draw[line width=1.0pt, draw=red] (axis cs:19.3472354443452,-3.99308131128253) rectangle (axis cs:22,3.93068941579374);
\end{axis}

\end{tikzpicture}%

%% file: tikz_plots/rd_roi/rd_roi3.tex
%
%
\begin{tikzpicture}
	
	\begin{axis}[%
		scale=0.2,
		width=11.04in,
		height=4.4in,
		at={(1.595in,4.769in)},
		scale only axis,
		point meta min=-30,
		point meta max=10,
		axis on top,
		xmin=0.550531089879742,
		xmax=79.9843026296711,
		tick align=outside,
		xlabel style={font=\color{white!15!black}},
		xlabel={Range in m},
		ymin=-3.99308131128253,
		ymax=3.93068941579374,
		ylabel style={font=\color{white!15!black}},
		ylabel={Velocity in m/s},
		axis background/.style={fill=white},
		xminorgrids,
		yminorgrids,
		xmin=12,
		xmax=42,
		ymin=-3.9,
		ymax=3.9,
		ytick={-3.9,-2,0,2,3.9},
		yticklabels={-4,-2,0,2,4},
		ylabel style={font=\color{white!15!black}, yshift=-0.34cm},
		ylabel={Velocity in m/s},
		colormap={mymap}{[1pt] rgb(0pt)=(0.2422,0.1504,0.6603); rgb(1pt)=(0.2444,0.1534,0.6728); rgb(2pt)=(0.2464,0.1569,0.6847); rgb(3pt)=(0.2484,0.1607,0.6961); rgb(4pt)=(0.2503,0.1648,0.7071); rgb(5pt)=(0.2522,0.1689,0.7179); rgb(6pt)=(0.254,0.1732,0.7286); rgb(7pt)=(0.2558,0.1773,0.7393); rgb(8pt)=(0.2576,0.1814,0.7501); rgb(9pt)=(0.2594,0.1854,0.761); rgb(11pt)=(0.2628,0.1932,0.7828); rgb(12pt)=(0.2645,0.1972,0.7937); rgb(13pt)=(0.2661,0.2011,0.8043); rgb(14pt)=(0.2676,0.2052,0.8148); rgb(15pt)=(0.2691,0.2094,0.8249); rgb(16pt)=(0.2704,0.2138,0.8346); rgb(17pt)=(0.2717,0.2184,0.8439); rgb(18pt)=(0.2729,0.2231,0.8528); rgb(19pt)=(0.274,0.228,0.8612); rgb(20pt)=(0.2749,0.233,0.8692); rgb(21pt)=(0.2758,0.2382,0.8767); rgb(22pt)=(0.2766,0.2435,0.884); rgb(23pt)=(0.2774,0.2489,0.8908); rgb(24pt)=(0.2781,0.2543,0.8973); rgb(25pt)=(0.2788,0.2598,0.9035); rgb(26pt)=(0.2794,0.2653,0.9094); rgb(27pt)=(0.2798,0.2708,0.915); rgb(28pt)=(0.2802,0.2764,0.9204); rgb(29pt)=(0.2806,0.2819,0.9255); rgb(30pt)=(0.2809,0.2875,0.9305); rgb(31pt)=(0.2811,0.293,0.9352); rgb(32pt)=(0.2813,0.2985,0.9397); rgb(33pt)=(0.2814,0.304,0.9441); rgb(34pt)=(0.2814,0.3095,0.9483); rgb(35pt)=(0.2813,0.315,0.9524); rgb(36pt)=(0.2811,0.3204,0.9563); rgb(37pt)=(0.2809,0.3259,0.96); rgb(38pt)=(0.2807,0.3313,0.9636); rgb(39pt)=(0.2803,0.3367,0.967); rgb(40pt)=(0.2798,0.3421,0.9702); rgb(41pt)=(0.2791,0.3475,0.9733); rgb(42pt)=(0.2784,0.3529,0.9763); rgb(43pt)=(0.2776,0.3583,0.9791); rgb(44pt)=(0.2766,0.3638,0.9817); rgb(45pt)=(0.2754,0.3693,0.984); rgb(46pt)=(0.2741,0.3748,0.9862); rgb(47pt)=(0.2726,0.3804,0.9881); rgb(48pt)=(0.271,0.386,0.9898); rgb(49pt)=(0.2691,0.3916,0.9912); rgb(50pt)=(0.267,0.3973,0.9924); rgb(51pt)=(0.2647,0.403,0.9935); rgb(52pt)=(0.2621,0.4088,0.9946); rgb(53pt)=(0.2591,0.4145,0.9955); rgb(54pt)=(0.2556,0.4203,0.9965); rgb(55pt)=(0.2517,0.4261,0.9974); rgb(56pt)=(0.2473,0.4319,0.9983); rgb(57pt)=(0.2424,0.4378,0.9991); rgb(58pt)=(0.2369,0.4437,0.9996); rgb(59pt)=(0.2311,0.4497,0.9995); rgb(60pt)=(0.225,0.4559,0.9985); rgb(61pt)=(0.2189,0.462,0.9968); rgb(62pt)=(0.2128,0.4682,0.9948); rgb(63pt)=(0.2066,0.4743,0.9926); rgb(64pt)=(0.2006,0.4803,0.9906); rgb(65pt)=(0.195,0.4861,0.9887); rgb(66pt)=(0.1903,0.4919,0.9867); rgb(67pt)=(0.1869,0.4975,0.9844); rgb(68pt)=(0.1847,0.503,0.9819); rgb(69pt)=(0.1831,0.5084,0.9793); rgb(70pt)=(0.1818,0.5138,0.9766); rgb(71pt)=(0.1806,0.5191,0.9738); rgb(72pt)=(0.1795,0.5244,0.9709); rgb(73pt)=(0.1785,0.5296,0.9677); rgb(74pt)=(0.1778,0.5349,0.9641); rgb(75pt)=(0.1773,0.5401,0.9602); rgb(76pt)=(0.1768,0.5452,0.956); rgb(77pt)=(0.1764,0.5504,0.9516); rgb(78pt)=(0.1755,0.5554,0.9473); rgb(79pt)=(0.174,0.5605,0.9432); rgb(80pt)=(0.1716,0.5655,0.9393); rgb(81pt)=(0.1686,0.5705,0.9357); rgb(82pt)=(0.1649,0.5755,0.9323); rgb(83pt)=(0.161,0.5805,0.9289); rgb(84pt)=(0.1573,0.5854,0.9254); rgb(85pt)=(0.154,0.5902,0.9218); rgb(86pt)=(0.1513,0.595,0.9182); rgb(87pt)=(0.1492,0.5997,0.9147); rgb(88pt)=(0.1475,0.6043,0.9113); rgb(89pt)=(0.1461,0.6089,0.908); rgb(90pt)=(0.1446,0.6135,0.905); rgb(91pt)=(0.1429,0.618,0.9022); rgb(92pt)=(0.1408,0.6226,0.8998); rgb(93pt)=(0.1383,0.6272,0.8975); rgb(94pt)=(0.1354,0.6317,0.8953); rgb(95pt)=(0.1321,0.6363,0.8932); rgb(96pt)=(0.1288,0.6408,0.891); rgb(97pt)=(0.1253,0.6453,0.8887); rgb(98pt)=(0.1219,0.6497,0.8862); rgb(99pt)=(0.1185,0.6541,0.8834); rgb(100pt)=(0.1152,0.6584,0.8804); rgb(101pt)=(0.1119,0.6627,0.877); rgb(102pt)=(0.1085,0.6669,0.8734); rgb(103pt)=(0.1048,0.671,0.8695); rgb(104pt)=(0.1009,0.675,0.8653); rgb(105pt)=(0.0964,0.6789,0.8609); rgb(106pt)=(0.0914,0.6828,0.8562); rgb(107pt)=(0.0855,0.6865,0.8513); rgb(108pt)=(0.0789,0.6902,0.8462); rgb(109pt)=(0.0713,0.6938,0.8409); rgb(110pt)=(0.0628,0.6972,0.8355); rgb(111pt)=(0.0535,0.7006,0.8299); rgb(112pt)=(0.0433,0.7039,0.8242); rgb(113pt)=(0.0328,0.7071,0.8183); rgb(114pt)=(0.0234,0.7103,0.8124); rgb(115pt)=(0.0155,0.7133,0.8064); rgb(116pt)=(0.0091,0.7163,0.8003); rgb(117pt)=(0.0046,0.7192,0.7941); rgb(118pt)=(0.0019,0.722,0.7878); rgb(119pt)=(0.0009,0.7248,0.7815); rgb(120pt)=(0.0018,0.7275,0.7752); rgb(121pt)=(0.0046,0.7301,0.7688); rgb(122pt)=(0.0094,0.7327,0.7623); rgb(123pt)=(0.0162,0.7352,0.7558); rgb(124pt)=(0.0253,0.7376,0.7492); rgb(125pt)=(0.0369,0.74,0.7426); rgb(126pt)=(0.0504,0.7423,0.7359); rgb(127pt)=(0.0638,0.7446,0.7292); rgb(128pt)=(0.077,0.7468,0.7224); rgb(129pt)=(0.0899,0.7489,0.7156); rgb(130pt)=(0.1023,0.751,0.7088); rgb(131pt)=(0.1141,0.7531,0.7019); rgb(132pt)=(0.1252,0.7552,0.695); rgb(133pt)=(0.1354,0.7572,0.6881); rgb(134pt)=(0.1448,0.7593,0.6812); rgb(135pt)=(0.1532,0.7614,0.6741); rgb(136pt)=(0.1609,0.7635,0.6671); rgb(137pt)=(0.1678,0.7656,0.6599); rgb(138pt)=(0.1741,0.7678,0.6527); rgb(139pt)=(0.1799,0.7699,0.6454); rgb(140pt)=(0.1853,0.7721,0.6379); rgb(141pt)=(0.1905,0.7743,0.6303); rgb(142pt)=(0.1954,0.7765,0.6225); rgb(143pt)=(0.2003,0.7787,0.6146); rgb(144pt)=(0.2061,0.7808,0.6065); rgb(145pt)=(0.2118,0.7828,0.5983); rgb(146pt)=(0.2178,0.7849,0.5899); rgb(147pt)=(0.2244,0.7869,0.5813); rgb(148pt)=(0.2318,0.7887,0.5725); rgb(149pt)=(0.2401,0.7905,0.5636); rgb(150pt)=(0.2491,0.7922,0.5546); rgb(151pt)=(0.2589,0.7937,0.5454); rgb(152pt)=(0.2695,0.7951,0.536); rgb(153pt)=(0.2809,0.7964,0.5266); rgb(154pt)=(0.2929,0.7975,0.517); rgb(155pt)=(0.3052,0.7985,0.5074); rgb(156pt)=(0.3176,0.7994,0.4975); rgb(157pt)=(0.3301,0.8002,0.4876); rgb(158pt)=(0.3424,0.8009,0.4774); rgb(159pt)=(0.3548,0.8016,0.4669); rgb(160pt)=(0.3671,0.8021,0.4563); rgb(161pt)=(0.3795,0.8026,0.4454); rgb(162pt)=(0.3921,0.8029,0.4344); rgb(163pt)=(0.405,0.8031,0.4233); rgb(164pt)=(0.4184,0.803,0.4122); rgb(165pt)=(0.4322,0.8028,0.4013); rgb(166pt)=(0.4463,0.8024,0.3904); rgb(167pt)=(0.4608,0.8018,0.3797); rgb(168pt)=(0.4753,0.8011,0.3691); rgb(169pt)=(0.4899,0.8002,0.3586); rgb(170pt)=(0.5044,0.7993,0.348); rgb(171pt)=(0.5187,0.7982,0.3374); rgb(172pt)=(0.5329,0.797,0.3267); rgb(173pt)=(0.547,0.7957,0.3159); rgb(175pt)=(0.5748,0.7929,0.2941); rgb(176pt)=(0.5886,0.7913,0.2833); rgb(177pt)=(0.6024,0.7896,0.2726); rgb(178pt)=(0.6161,0.7878,0.2622); rgb(179pt)=(0.6297,0.7859,0.2521); rgb(180pt)=(0.6433,0.7839,0.2423); rgb(181pt)=(0.6567,0.7818,0.2329); rgb(182pt)=(0.6701,0.7796,0.2239); rgb(183pt)=(0.6833,0.7773,0.2155); rgb(184pt)=(0.6963,0.775,0.2075); rgb(185pt)=(0.7091,0.7727,0.1998); rgb(186pt)=(0.7218,0.7703,0.1924); rgb(187pt)=(0.7344,0.7679,0.1852); rgb(188pt)=(0.7468,0.7654,0.1782); rgb(189pt)=(0.759,0.7629,0.1717); rgb(190pt)=(0.771,0.7604,0.1658); rgb(191pt)=(0.7829,0.7579,0.1608); rgb(192pt)=(0.7945,0.7554,0.157); rgb(193pt)=(0.806,0.7529,0.1546); rgb(194pt)=(0.8172,0.7505,0.1535); rgb(195pt)=(0.8281,0.7481,0.1536); rgb(196pt)=(0.8389,0.7457,0.1546); rgb(197pt)=(0.8495,0.7435,0.1564); rgb(198pt)=(0.86,0.7413,0.1587); rgb(199pt)=(0.8703,0.7392,0.1615); rgb(200pt)=(0.8804,0.7372,0.165); rgb(201pt)=(0.8903,0.7353,0.1695); rgb(202pt)=(0.9,0.7336,0.1749); rgb(203pt)=(0.9093,0.7321,0.1815); rgb(204pt)=(0.9184,0.7308,0.189); rgb(205pt)=(0.9272,0.7298,0.1973); rgb(206pt)=(0.9357,0.729,0.2061); rgb(207pt)=(0.944,0.7285,0.2151); rgb(208pt)=(0.9523,0.7284,0.2237); rgb(209pt)=(0.9606,0.7285,0.2312); rgb(210pt)=(0.9689,0.7292,0.2373); rgb(211pt)=(0.977,0.7304,0.2418); rgb(212pt)=(0.9842,0.733,0.2446); rgb(213pt)=(0.99,0.7365,0.2429); rgb(214pt)=(0.9946,0.7407,0.2394); rgb(215pt)=(0.9966,0.7458,0.2351); rgb(216pt)=(0.9971,0.7513,0.2309); rgb(217pt)=(0.9972,0.7569,0.2267); rgb(218pt)=(0.9971,0.7626,0.2224); rgb(219pt)=(0.9969,0.7683,0.2181); rgb(220pt)=(0.9966,0.774,0.2138); rgb(221pt)=(0.9962,0.7798,0.2095); rgb(222pt)=(0.9957,0.7856,0.2053); rgb(223pt)=(0.9949,0.7915,0.2012); rgb(224pt)=(0.9938,0.7974,0.1974); rgb(225pt)=(0.9923,0.8034,0.1939); rgb(226pt)=(0.9906,0.8095,0.1906); rgb(227pt)=(0.9885,0.8156,0.1875); rgb(228pt)=(0.9861,0.8218,0.1846); rgb(229pt)=(0.9835,0.828,0.1817); rgb(230pt)=(0.9807,0.8342,0.1787); rgb(231pt)=(0.9778,0.8404,0.1757); rgb(232pt)=(0.9748,0.8467,0.1726); rgb(233pt)=(0.972,0.8529,0.1695); rgb(234pt)=(0.9694,0.8591,0.1665); rgb(235pt)=(0.9671,0.8654,0.1636); rgb(236pt)=(0.9651,0.8716,0.1608); rgb(237pt)=(0.9634,0.8778,0.1582); rgb(238pt)=(0.9619,0.884,0.1557); rgb(239pt)=(0.9608,0.8902,0.1532); rgb(240pt)=(0.9601,0.8963,0.1507); rgb(241pt)=(0.9596,0.9023,0.148); rgb(242pt)=(0.9595,0.9084,0.145); rgb(243pt)=(0.9597,0.9143,0.1418); rgb(244pt)=(0.9601,0.9203,0.1382); rgb(245pt)=(0.9608,0.9262,0.1344); rgb(246pt)=(0.9618,0.932,0.1304); rgb(247pt)=(0.9629,0.9379,0.1261); rgb(248pt)=(0.9642,0.9437,0.1216); rgb(249pt)=(0.9657,0.9494,0.1168); rgb(250pt)=(0.9674,0.9552,0.1116); rgb(251pt)=(0.9692,0.9609,0.1061); rgb(252pt)=(0.9711,0.9667,0.1001); rgb(253pt)=(0.973,0.9724,0.0938); rgb(254pt)=(0.9749,0.9782,0.0872); rgb(255pt)=(0.9769,0.9839,0.0805)},
		colorbar,
		colorbar style={ytick={-26.2,-8.2,9.8}, yticklabels={-40,-20,0}, style={xshift=-0.2cm}},
		]
		\addplot [forget plot] graphics [xmin=0.550531089879742, xmax=79.9843026296711, ymin=-3.99308131128253, ymax=3.93068941579374] {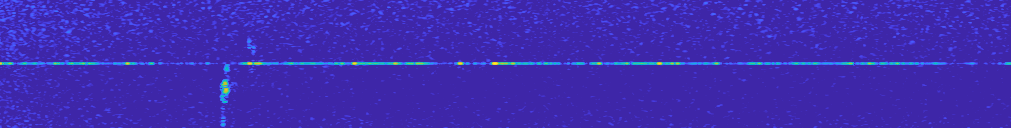};
		\draw[line width=1.0pt, draw=red] (axis cs:17,-3.99308131128253) rectangle (axis cs:19.0774760125387,3.93068941579374);
	\end{axis}
\end{tikzpicture}%